\documentclass{aa}  

\usepackage{graphicx}
\usepackage{threeparttable}
\usepackage[varg]{txfonts}
\usepackage{appendix}
\bibpunct{(}{)}{;}{a}{}{,}
\usepackage{color}

\begin{document}

\title{ Precise Masses, Ages of $\sim$1.0 million  RGB and RC stars observed by the LAMOST \footnote{The catalogue will be available at CDS via https://cdsarc.cds.unistra.fr/cgi-bin/qcat?J/A+A/ and the LAMOST official website via http://www.lamost.org/dr8/v1.0/doc/vac.}}

\author{Chun Wang
	\inst{1}
          \and
         Yang Huang
         \inst{2,3}
         \and
         Yutao Zhou 
         \inst{4}
         \and
         Huawei Zhang
         \inst{5,6}
         }

\institute{Tianjin Astrophysics Center, Tianjin Normal University, Tianjin 300387,  China; 
              \email{wchun@tjnu.edu.cn}
         \and
         University of Chinese Academy of Sciences, Beijing 100049,  China;  
         \email{huangyang@bao.ac.cn}
         \and
         National Astronomical Observatories, Chinese Academy of Sciences, Beijing 100012,  China
        \and
        College of Mathematics and Physics, Guangxi Minzu University, Nanning 530006, China
        	\and
	Department of Astronomy, School of Physics, Peking Univeristy, Beijing 100871,  China
	\and
	Kavli Institute for Astronomy and Astrophysics, Peking University, Beijing 100871, China
        }

\date{Received ---; accepted ---}

%\pagerange{\pageref{firstpage}--\pageref{lastpage}} \pubyear{except for 2022}

%\maketitle

\abstract{We construct a catalogue of stellar masses and ages for 696,680  red giant branch (RGB) stars, 180,436   primary red clump (RC) stars, and 120,907  secondary RC stars selected from the LAMOST\,DR8.  The RGBs, primary RCs, and secondary RCs are identified with the large frequency spacing ($\Delta \nu$) and period spacing ($\Delta P$), estimated from the LAMOST spectra with spectral SNRs $> 10$ by the neural network method supervised with the seismologic information from LAMOST-Kepler sample stars. The purity and completeness of both RGB and RC samples are better than 95\% and 90\%, respectively.  The mass and age of RGBs and RCs are determined again with the neural network method by taking the LAMOST-Kepler giant stars as the training set.  The typical uncertainties of stellar mass and age are, respectively, 10\% and 30\% for the RGB stellar sample.  For RCs,  the typical uncertainties of stellar mass and age are  9\% and 24\%, respectively.   The RGB and RC stellar samples cover a large volume of the Milky Way (5 $< R < 20$\,kpc and $|Z| <$\,5\,kpc), which are valuable data sets for various Galactic studies. }

\keywords
{Stars: fundamental parameters}

\titlerunning{Masses and Ages of red giant stars}
\authorrunning{C. Wang}

\maketitle

\section{Introduction}
The Milky Way (MW) is an excellent laboratory to test galaxy formation and evolution histories, because it is the only galaxy in which the individual stars can be resolved in exquisite detail.   One can study the properties of the stellar population and assemblage history of the Galaxy using multidimensional information, including but not limited to accurate age, mass, stellar atmospheric parameters, chemical abundances, 3D positions, 3D  velocities.  Thanks to a number of large-scale spectroscopic surveys, e.g., the RAVE \citep{Steinmetz2006}, the SEGUE \citep{yanny-segue}, the LAMOST Experiment for Galactic Understanding and Exploration \citep[LEGUE; ][]{deng-legue,liu-lss-gac, Zhao2012}, the Galactic Archaeology with HERMES \citep[GALAH; ][]{DeSilva2015}, the Apache Point Observatory Galactic Evolution Experiment \citep[APOGEE; ][]{Majewski2017} and the Gaia missions \citep{gaiamission},  accurate stellar atmospheric parameters, chemical element abundance ratios, 3D positions and 3D  velocities for a huge sample of Galactic stars now is available.  

The age of stars can hardly be  measured directly  but can be 
generally derived indirectly from the photometric and spectroscopic data in combination with the stellar evolutionary models  \citep{Soderblom2010}.  Many methods have been explored to estimate the ages of stars at different evolutionary stages.  Matching with stellar isochrones by stellar atmospheric parameters yielded by spectra is adopted to estimate the ages for main-sequence turnoff and subgiant
stars \citep[e.g.,][]{Xiang2017, Sanders2018}, and the uncertainties of the estimated age are $\sim$\,20\%--30\%. 
%The evolved stars (RGB and RC stars) of different ages are tightly crowded together in the stellar isochrones. It is difficult to estimate their ages using isochrone fitting method.  
It is not easy to infer the age information for evolved red giant stars by the isochrone fitting method due to the mixing of stars of various evolutionary stages on the Herzsprung–Russell (HR) diagram.  
The degeneracy can be broken if the stellar mass is determined independently. 
 In this manner, age estimates with a precision of about 10\%--20\% have been achieved for subgiants \citep {Li2020}, red giants \citep{Tayar2017, Wu2018, Bellinger2020, Lit2022, Wu2023} and red clump stars \citep{Huang2020} in the Kepler fields, since these stars have precise asteroseismic masses with uncertainties of 8\%--10\% \citep[e.g.,][]{Huber2014, Yu2018}.  Those asteroseismic masses and ages are adopted as training data to estimate the masses and ages directly from stellar spectra using a data-driven method \citep[e.g.][]{Ness2016, Ho2017, Ting2018, Wuyaqian2019, Huang2020}.  
 The physics behind the estimates of spectroscopic mass from the data-driven method is that stellar spectra carry key features to determine the carbon to nitrogen abundance ratio [C/N], which  is tightly
correlated with stellar mass as the result of the convective mixing through the CNO cycle (e.g., the first-dredge up process). 
Of course,  [C/N] ratios deducible from the optical/infrared spectra can be directly used to derive their ages using the relation between the regressed trained relation between [C/N] and masses \citep[e.g.,][]{Martig2016a, Ness2016}.   

By March 2021, the   Low-Resolution Spectroscopic Survey of LAMOST\,DR8 has released 11,214,076 optical (3700--9000\,\AA) spectra with a resolving power of $R$$\sim$1800, of which more than 90 per\,cent are stellar spectra. The classifications and radial velocity $V_{r}$ measurements for these spectra are provided by the official catalogue of LAMOST\,DR8 \citep{luo2015}.  The accurate stellar parameters, including the atmospheric stellar parameters, chemical element abundance ratios,  14 bands absolute magnitudes and photometric distance, have been provided by \cite{Wang2022}.

In this work, we have applied a similar technique used by \cite{Ting2018} to separate red giant branch (RGB) stars and red clump (RC) stars. The latter can be further divided into primary  RC stars and secondary RC stars, according to their stellar masses\footnote{The primary RC stars are ignited degenerately as the descendants of low-mass stars (typically smaller than 2\,$M_{\odot}$), while the secondary RC stars (ignited non-degenerately)  are the descendants of high-mass stars (typically larger than 2\,$M_{\odot}$).} \citep[e.g., ][]{Bovy2014, Huang2020}.  The ages and masses of these separated RGBs, primary RCs and secondary RCs will be estimated using the neural network machine learning technique by adopting  LAMOST--Kepler common stars as the training sample.  Other stellar parameters estimated by  \cite{Wang2022}, together with the information from the LAMOST DR8  and the astrometric information from Gaia EDR3 \citep{Gaiaedr3}, are also provided.

The paper is organized as follows.  Section\,2 introduces the data employed in this work. Section\,3 presents the selections of RGBs, primary RCs and secondary RCs. In Section\,4, we present estimates of ages and masses derived from  LAMOST spectra using the neural network technique.  A detailed uncertainty analysis is presented in Section\,5. In Section\,6, we introduce the properties of our RGB, primary RC and secondary RC stellar samples. Finally, it is a summary in Section\,7.   
\section{Data}
The LAMOST Galactic survey \citep{deng-legue,liu-lss-gac, Zhao2012} is a spectroscopic survey to obtain over 10 million stellar spectra.  11,214,076 optical (3700--9000\,\AA) low-resolution spectra ($R$$\sim$1800) have been released by March 2021, of which more than 90 per\,cent are stellar spectra. 

The value-added catalogue of LAMOST\,DR8 by \cite{Wang2022} provide stellar atmospheric parameters ($T_{\mathrm{eff}}$, $\log\,g$,  $ \mathrm{ \, [Fe/H]} $ and [M/H]), chemical elemental abundance to metal or iron ratios ([$\alpha$/M], [C/Fe], [N/Fe]), absolute magnitudes of 14 photometric bands, i.e., $G, Bp, Rp$ of Gaia, $J, H, K\rm s$ of 2MASS, $W1, W2$ of WISE, $B, V, r$ of APASS and $g, r, i$ of SDSS,  and spectro-photometric distances for   4.9 million unique stars targeted by   LAMOST\,DR8.  
This value-added catalogue of LAMOST DR8 is publicly available at the CDS and the LAMOST official website \footnote{The catalogue will be available at  the LAMOST official website via http://www.lamost.org/dr8/v1.0/doc/vac.}.
For stars with spectral signal-to-noise ratios (SNRs) larger than 50, precisions of  $T_{\mathrm{eff}}$,  $\log g$,  [Fe/H], [M/H], [C/Fe], [N/Fe] and [$\alpha$/M] are 85\,K, 0.10\,dex, 0.05\,dex, 0.05\,dex, 0.05\,dex, 0.08\,dex and 0.03\,dex, respectively.  The typical uncertainties of 14  band's absolute magnitudes are 0.16--0.22\,mag for stars with spectral SNRs larger than 50, corresponding to a typical distance uncertainty of around 10\,\%.   This stellar sample in the catalogue 
contains main-sequence stars, main-sequence turn-off stars, sub-giant stars and giant stars of all evolutionary stages (e.g. RGB, RC, AGB etc.).  We can select samples of RGB and RC  from the catalogue as the first step.

\section{Selection of RGBs and RCs }
%\subsection{Selection of RGBs and RCs}

Before separating RGBs and RCs, we first select giant stars using the  $T_{\rm eff}$ and $\log g$ provided by \cite{Wang2022}. 
By cuts of $T_{\rm eff} \le 5800$\,K and log\,$g \le$\,3.8, a total of about 1.3 million giant stars are selected.   
 RGB and RC are stars with burning hydrogen in a shell around an inert helium core and stars with helium-core and hydrogen-shell burning, respectively.  However,  the RGBs and RCs occupy overlapping parameter spaces in a classical Herzsprung–Russell diagram (HRD). This is due to the fact that the RGBs and RCs can have quite similar surface characteristics such as effective temperature, surface gravity and luminosity \citep{Elsworth2017, Wuyaqian2019}. The classical method of classifying  RGB and RC is based on the $T_{\rm eff}$--log\,$g$--[Fe/H] relation as well as colour-metallicity diagram. The typical contamination rate of RC from RGB stars using this method could be better  than $\sim$10\% \citep{Bovy2014, Huang2015, Huang2020}.  
%Thus  it is difficult to  distinguish RGB and RC stars  \textbf{with similar parameters} in the classical HRD. 

%\textbf{On the basis of the $T_{\rm eff}$--log\,$g$--[Fe/H]  and  color-magnitude relations, the  full spectral information can provide better classifications of RC and RGB \citep{Ting2018,Wuyaqian2019}.}  
Now, asteroseismology has become the gold standard for separating RGBs and RCs \citep{Montalb2010, Bedding2011, Mosser2011, Mosser2012, Vrard2016, Hawkins2018}.
Although RGBs and RCs have similar surface characteristics, they have totally different interior structures.  The solar-like oscillations in red giant stars are excited and intrinsically damped by turbulence in the 
envelopes near-surface convection and can have acoustic (p-mode) and gravity (g-mode) characteristics \citep{Chaplin2013, Wuyaqian2019}.   The p-mode and g-mode are always associated with the stellar envelope with pressure as the restoring force and the inner core with buoyancy as the restoring force, respectively. 
Mixed modes character also exists, displaying g-mode-like behaviour in the central region of a star, and p-mode-like behaviour in the envelope. The evolved red giant stars always show mixed mode.  The core density of RCs is lower than that of RGBs with the same luminosity, which causes a significantly stronger coupling between g- and p-modes and leads to larger period spacing \citep{Bedding2011, Wuyaqian2019}.  Thus one can distinguish RGBs and RCs from the period spacing ($\Delta P$).  There are two kinds of RCs, the primary RC stars (formed
from lower-mass stars) and the secondary RC stars (formed from more massive stars). The secondary RC stars have larger $\Delta \nu$ compared to primary RC stars.  Thus, the asteroseismology is a powerful tool to separate RGBs, primary RCs and secondary RCs \citep{Bedding2011, Stello2013, Pinsonneault2014, Vrard2016, Elsworth2017, Wuyaqian2019}, the typical contamination rate of RC from RGB using the method is only 3\% \citep{Ting2018, Wuyaqian2019}.  

The photospheric abundances reflect the interior structure of stars via the efficacy of extra-mixing on the upper RGB\citep{Martell2008, Masseron2015, Masseron2017a, Masseron2017b, Ting2018}. Thus, one can directly derive $\Delta \nu$ and $\Delta P$ from low-resolution spectra with a data-driven method using stars with accurate estimates of $\Delta \nu$ and $\Delta P$ \citep{Ting2018, Wuyaqian2019} as training stars and finally separate RGBs, primary RCs and secondary RCs.

The $\Delta \nu$ and $\Delta P$ of about  6100 Kepler giants are estimated through the Fourier analysis of their light curves by \cite{Vrard2016}.  Amongst them, 2662 unique stars have a good quality of LAMOST spectrum (S/N$\,> 50$), including 826 RGBs, 1599 primary RCs and 237 secondary RCs, which is a good training sample for deriving $\Delta \nu$ and $\Delta P$ from LAMOST spectra.  Here, we select 1800 stars as training stars to estimate  $\Delta \nu$ and $\Delta P$ from LAMOST spectra. Other stars are selected as testing stars. First, we pre-process LAMOST spectra and build up a neural network model to construct the relation between the pre-processed LAMOST spectra and asteroseismic parameters, i.e.,   $\Delta \nu$ and $\Delta P$.  Then we estimate  $\Delta \nu$ and $\Delta P$ for all testing stars using LAMOST spectra based on the neural network model.  The pre-processing of LAMOST spectra and neural network model are the same as that of \cite{Wang2022}.  As discussed in \cite{Wang2022}, we only use spectra in the wavelength range of 3900-6800\,\AA\,and 8450--8950\,\AA,  because of the low-quality spectra in the 3700--3900\,\AA\, for most of the stars and serious background contamination (including sky emission lines and telluric bands) and very limited effective information in the spectra of 6800--8450\,\AA. 

%Here we estimate $\Delta \nu$ and $\Delta P$ with the method  of neural network, which is similar with that of \cite{Wang2022}   

Following \cite{Wang2022}, a neural network contains three layers is build up, which can be written as:
\begin{equation}
P=\omega\sigma(\omega^{'}_{i}\sigma(\omega^{''}_{j}\sigma(\omega_{\lambda_{k}}f_{\lambda}+b_{k})+b_{j})+b^{'}_{i}),
\end{equation}
where $P$ is the asteroseismic parameters $\Delta \nu$ or $\Delta P$;  $\sigma$ is the Relu activation function; $\omega$ and b are
weights and biases of the network to be optimized; the index
$i$, $j$ and $k$ denote the number of neurons in the third, second and first layer; and $\lambda$ denotes the wavelength pixel. The neurons for the first, second and third layers are respectively  512, 256 and 64.  
The architecture of the neural network is empirically adjusted based on the performance of both training and testing samples. A delicate balance must be struck between achieving high accuracy of parameters and avoiding overfitting the training data.
The training process is carried out with the $Tensorflow$ package in $Python$. 

\begin{figure*}
\centering
\includegraphics[width=6.0in]{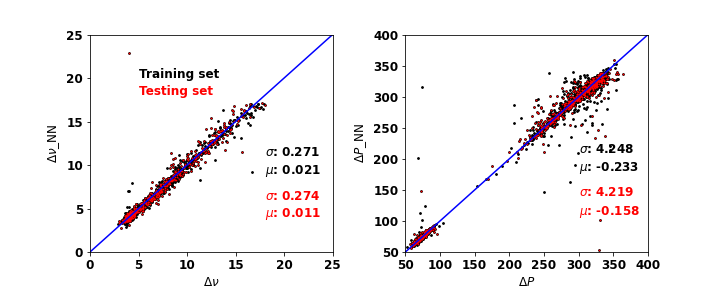}
\caption{The comparisons of $\Delta \nu$ and $\Delta P$ provided by \citet{Vrard2016} ($X$-axis) using Kepler data and that derived by neural network model ($Y$-axis) for $1,800$ training (black dots) and $862$ testing stars (red dots). The values of the mean and standard deviation of the differences are labelled in the bottom-right corner of each panel. }
\label{compare_test_dnu_dp}
\end{figure*}

After estimating asteroseismic parameters for training and testing samples, we compare our frequency spacing and period spacing ($\Delta \nu_{\rm NN}$ and $\Delta P_{\rm NN}$) with these provided by \cite{Vrard2016}  for testing stars, as shown in Fig.\,\ref{compare_test_dnu_dp}. 
  We can find that our $\Delta \nu_{\rm NN}$ and $\Delta P_{\rm NN}$ match well with these of \cite{Vrard2016}. Our neural network could derive precise frequency spacing and period spacing from LAMOST spectra.

\begin{figure*}
\centering
\includegraphics[width=6.0in]{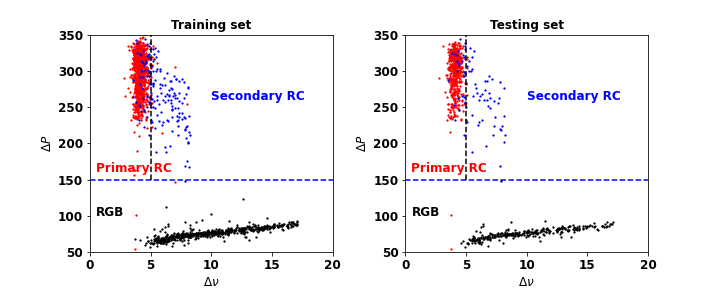}
\caption{The stellar distributions on the plane of $\Delta \nu$-$\Delta P$ derived by neural network method for 1,800 training (left panel) and 862 testing stars (right panel). The black, red and blue dots show the 826 RGBs, 1,599 primary RCs and 237 secondary RCs classified by \citet{Vrard2016} using Kepler data, respectively. The blue dashed line ($\Delta P = 150$\,s) is plotted to separate RGBs and RCs. The black dashed line ($\Delta \nu = 5\mu$Hz) is plotted to separate primary RCs and secondary RCs. }
\label{select_rgb_rc_predict}
\end{figure*}

The $\Delta P$ of  RC and RGB  stars are very different, they are $\sim$ 300s and $\sim$ 70s, respectively \citep{Bedding2011, Ting2018}.  Fig.\,\ref{select_rgb_rc_predict} shows the stellar distributions on the plane of $\Delta \nu$ and $\Delta P$ derived by neural network method for training and testing sample.   The plots clearly show that a gap of $\Delta P$ between the RGB and RC stellar populations is visible for both training and testing samples. It illustrates that the spectroscopically estimated frequency spacing and period spacing ($\Delta \nu$ and $\Delta P$ ) using our neural network method, especially $\Delta P$,  could allow one to separate the RC stars from the RGB stars.   Here we identify giant stars with $\Delta P > 150$\,s as RCs, and giant stars with  $\Delta P \leq 150$\,s as red giant stars.

 Fig.\,\ref{select_rgb_rc_predict} also suggests that it is also possible to classify primary and secondary RC stars using the spectroscopically estimated frequency spacing and period spacing.  RC stars with $\Delta \nu > 6\mu$Hz can be classified as secondary RC stars \citep{Yang2012, Ting2018}.   To separate the primary RCs from the secondary RCs, we select $\Delta P > 150$\,s and $\Delta \nu > 5\mu$Hz as secondary RCs, $\Delta P > 150$\,s and $\Delta \nu \leq 5\mu$Hz as primary RCs.  
 %For the values of $\Delta \nu$ and $\Delta P$ provided by \cite{Vrard2016}, one can use the same criterion to distinguish RGBs, PRCs and SRCs clearly (one can see Fig.\,\ref{select_rgb_rc_vrard} in Appendix for more details). 

The completeness of RGB, primary RC, and secondary RC are respectively 99.2\%(280/282), 96.8\%(485/501), and 62\%(49/79) examined by the testing set. The  purity of RGB, primary RC, and secondary RC are respectively 98.9\% (280/283), 94.2\% (485/515), and 79\%
(49/62) examined again by the testing set. 

We adopt the constructed neural network model to $\sim$1.3 million LAMOST spectra collected by LAMOST DR8 of giant stars to estimate their spectroscopic frequency spacing and period spacing.   With the aforementioned criteria of separating RGBs, primary RCs and secondary RCs,  937,082, 244,458, and 167,105 spectra for 696,680 unique RGBs, 180,436 unique primary RCs and 120,907 unique secondary RCs with spectral SNRs $>10$ are found.  For a star with multiple observed spectra, the classification, mass and age derived by the spectrum with the highest spectral SNR are suggested. 

\section{Ages and masses of RGBs and RCs}
In this Section, we estimate the masses and ages of our RGBs, primary RCs and secondary RCs selected from the LAMOST low-resolution spectroscopic survey. Doing so, we first estimate the masses and ages using the respective asteroseismic information and isochrone fitting method,  for LAMOST-Kepler giant stars.  Then we estimate masses and ages for all LAMOST DR8 giant stars,  using the neural network method by taking the  LAMOST-Kepler giant stars as the training set. For both RGBs and RCs, we build up two different neural network models to estimate ages and masses, respectively.  

\subsection{Ages and masses of LAMOST-Kepler training and testing stars}
 Masses of RGBs and RCs can be accurately measured from the asteroseismic information achieved by Kepler mission.
 We cross-match our RGB and RC samples to the Kepler sample with accurate asteroseismic parameters measured by \cite{Yu2018}, using light curves of stars provided by the Kepler mission.  In total, 3,220 and  3,276 common RGBs and RCs are found with both high-quality spectra (SNR$>$50), well-derived spectroscopic and precise asteroseismic parameters.

\begin{figure*}
\centering
\includegraphics[width=6.0in]{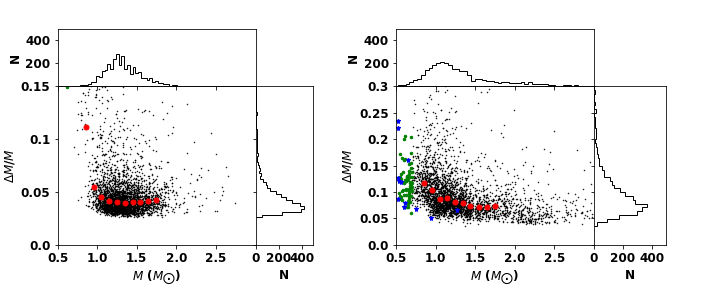}
\caption{Distributions of 3,220 LAMOST-Kpler common RGBs (left panels) and 3,276 LAMOST-Kpler common RCs (right panels) on the $M$--$\Delta M/M$ plane.  $M$ and $\Delta M$ are the estimated masses and their associated uncertainties. The red dot presents the median values in each mass bin with a bin size of 0.1\,$M_{\odot}$. Green dots are stars with $M < 0.7\,M_{\odot}$. Blue stars show the post-mass-transfer helium-burning stars identified by \cite{Li2022}.  Distributions of masses and their associated uncertainties are over-plotted using histograms.}
\label{mass_error}
\end{figure*}

For these RGB and RC common stars, we determine their asteroseismic masses using  modified scaling relations,
\begin{equation}
\frac{M}{M_{\odot}}=(\frac{\Delta \nu}{f_{\Delta \nu}\Delta \nu_{\odot}})^{-4}(\frac{\nu_{\rm max}}{\nu_{\rm max,\odot}})^{3}(\frac{T_{\rm eff}}{T_{\rm eff,\odot}})^{3/2},
\end{equation}
Here the Solar values of $T_{\rm eff,\odot}$, $\nu_{\rm max,\odot}$  and $\Delta \nu_{\odot}$ are  5777\,K, 3090\,$\mu$Hz and 135.1\,$\mu$Hz \citep{Huber2011}, respectively. Where $f_{\Delta \nu}$, the correction factor can be obtained for all stars using the publicly available code $Asfgrid^{11}$ provided by \cite{Sharma2016}.
With this modified scaling relation, the masses of these  3,220 and  3,276 RGB and RC common stars are derived with precise $\Delta \nu$ and $\nu_{\rm max}$ provided by \cite{Yu2018} and effective temperature $T_{\rm eff}$ taken from \cite{Wang2022}. Fig.\,\ref{mass_error} shows the masses distributions estimated from the asteroseismic and spectroscopic stellar atmospheric parameters, as well as their associated uncertainties. The plot clearly shows that the typical uncertainties of masses of RGBs and RCs are respectively 4.1 and 8.4 per\,cent.

There are several stars with a mass lower than 0.7\,$\,M_{\odot}$ as shown in Fig.\,\ref{mass_error}, especially for RCs.  Some of these stars may be post-mass-transfer helium-burning stars \citep{Li2022}, which may be in binary systems, suffered a mass-transfer history with a mass-loss of about 0.1--0.2\,$\,M_{\odot}$.  We make a crossmatch to the list of post-mass-transfer helium-burning red giants provided by \cite{Li2022} and find twelve common stars, nine of them are stars with $M < 0.7\,M_{\odot}$. In the current work, we did not consider such an extreme mass-loss case and thus the readers should be careful of the parameters of a star with a mass smaller than 0.7\,$M_{\odot}$.

%When the $f_{\Delta \nu}$ is equal to 1, the relation is standard scaling relation, which will produce systematic errors of 10 to 15 per\,cent in the derived masses for RGB and RC stars \citep[e.g.,][]{Huber2011, Viani2017}.  The systematic errors  which can be obtained for all stars using the publicly available code $Asfgrid^{11}$ provided by \cite{Sharma2016}.  }

%\begin{equation}
%\frac{M}{M_{\odot}}=(\frac{\Delta \nu}{\Delta \nu_{\odot}})^{-4}(\frac{\nu_{\rm max}}{\nu_{\rm max,\odot}})^{3}(\frac{T_{\rm eff}}{T_{\rm eff,\odot}})^{3/2}.
%\end{equation}
%Here the Solar values of $T_{\rm eff,\odot}$, $\nu_{\rm max,\odot}$  and $\Delta \nu_{\odot}$ are  5777\,K, 3090\,$\mu$Hz and 135.1\,$\mu$Hz \citep{Huber2011}, respectively.  As discussed in  previous studies \citep[e.g.,][]{Huber2011, Viani2017}, the standard scaling relation may induce systematic errors of 10 to 15 per\,cent in the derived masses for RGB  and RC stars.  Similar to \cite{Wu2018} and \cite{Huang2020}, a modified scaling relation from \cite{Sharma2016} to reduce the systematic errors is adopted when estimating the asteroseismic masses. The modified relation is

%where $f_{\Delta \nu}$, the correction factor, can be obtained for all stars using the publicly available code $Asfgrid^{11}$ provided by \cite{Sharma2016}. 

We further estimate the ages of these common stars based on the stellar isochrone fitting method using the aforementioned estimated masses and the effective temperatures, metallicities and surface gravities provided by  \cite{Wang2022}.  Similar to that of \cite{Huang2020}, we adopt the PARSEC isochrones calculated with a mass-loss parameter $\eta_{\rm Reimers} = 0.2$ \citep{Bressan2012} as the stellar evolution tracks.
Here, we adopt one Bayesian approach similar to \cite{Xiang2017}  and \cite{Huang2020} when doing isochrone fitting.  The mass $M$, effective temperatures $T_{\rm eff}$, metallicity [M/H] and surface gravity $\log g$ are the input constraints.  

\begin{figure*}
\centering
\includegraphics[width=6.0in]{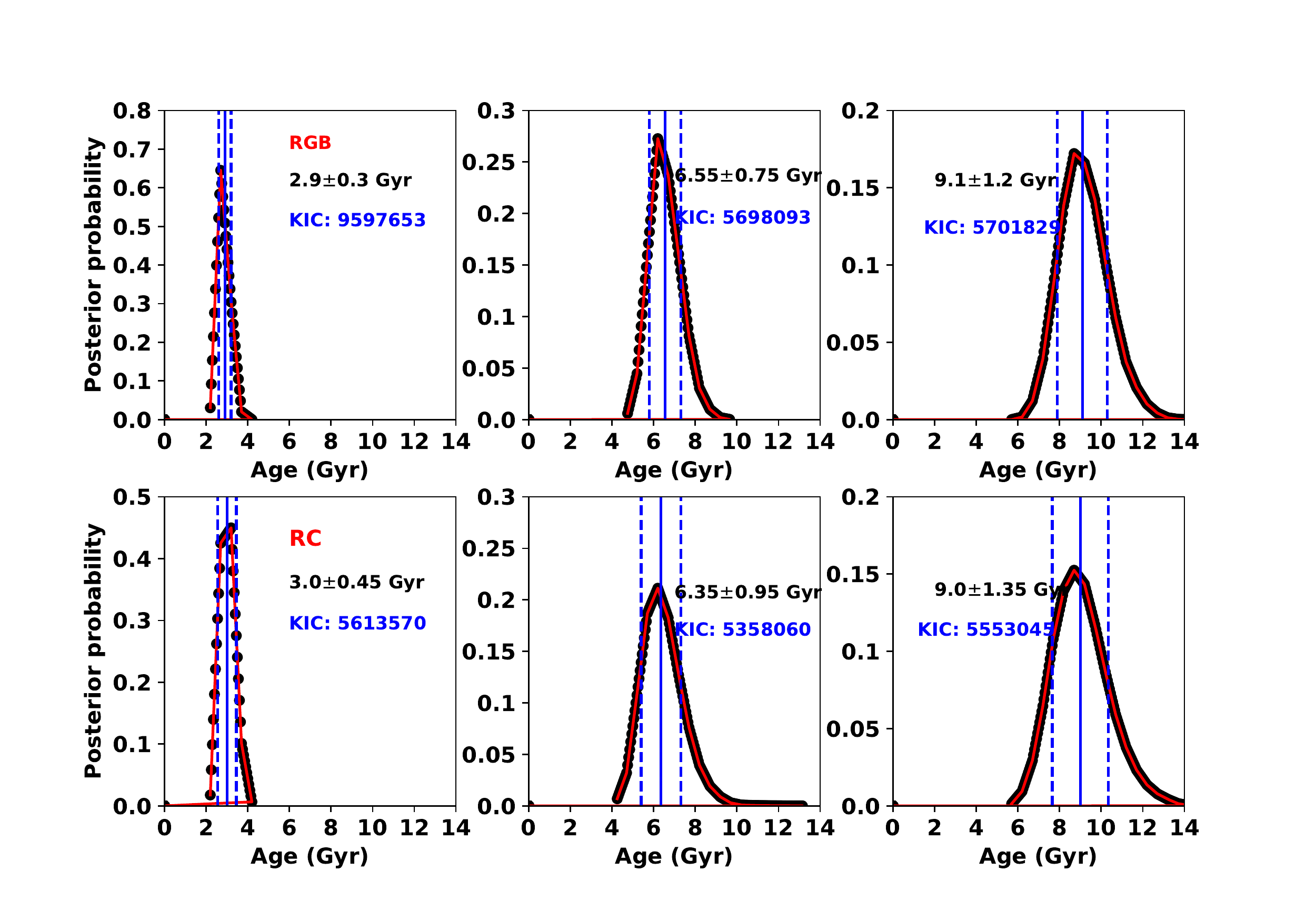}
\caption{The posterior probability distributions as a function of age for three stars of typical ages for both RGB (top panels) and RC (bottom panels) stars. The blue solid lines denote the median values of the stellar ages.  The blue dashed lines show the standard deviations of the stellar ages. The KIC number of the six stars are given here. }
\label{age_pdf}
\end{figure*}

\begin{figure*}
\centering
\includegraphics[width=6.0in]{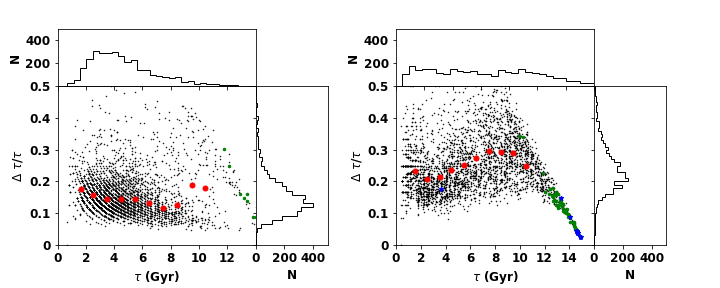}
\caption{Distributions of 3,220 LAMOST-Kepler common RGBs (left panels) and 3,276 LAMOST-Kepler common RCs (right panels) on the $\tau$--$\Delta \tau/\tau$ plane.  $\tau$ and $\Delta \tau$ are the estimated ages using the isochrone fitting method and their associated uncertainties. The red dot presents the median value in each age bin with a bin size of 1.0\,Gyr. Green dots are stars with $M < 0.7\,M_{\odot}$. Blue stars show stars the post-mass-transfer helium-burning stars identified by \cite{Li2022}.  Distributions of ages and their associated uncertainties are over-plotted using histograms. }
\label{age_error}
\end{figure*}

The posterior probability distributions as a function of age for three stars of typical ages for both RGB and RC stars are shown in Fig.\,\ref{age_pdf}.  As shown in Fig.\,\ref{age_pdf}, we can find that the age distributions show prominent peaks, which suggest that the resultant ages are well constrained, especially thanks to the precise mass estimates from asteroseismology information.  Fig.\,\ref{age_error} shows the distributions of ages,  as well as the uncertainties,  of RGBs and RCs,  estimated using the isochrone fitting method.  
Both the RGB and RC samples cover a whole range of possible ages of stars, from close to zero to 13\,Gyr (close to the universe age). The typical age uncertainties of  RGB and RC stars are 14 and 23 per\,cent, respectively.  There are only a small fraction stars that have age uncertainties larger than 40 per\,cent.  Larger age uncertainties of RCs compared to RGBs are the consequence of larger mass uncertainties of RCs. 
It is noted that the relationship between age and its associated uncertainty of RGBs is opposite to that of RCs at $\tau < 8$\,Gyr, which may be the consequence of the larger variations of mass uncertainty of RCs compared to that of RGBs at $1.0 < M < 1.5\,M_{\odot}$. 
The mass uncertainty decrease 1.7\% and 0.02\% when mass decrease from 1.55\,$\,M_{\odot}$ ($\tau \sim 2.0$\,Gyr) to 1.05\,$\,M_{\odot}$  ($\tau \sim 9.0$\,Gyr) for RCs and RGBs, respectively. 
Large mass uncertainties produce large age uncertainties ($\Delta \tau$), thus older  RGBs and RCs have respectively smaller and larger relative age uncertainties ($\Delta \tau/\tau$).  %The relative age error decreases as age increases for RCs at $\tau > 10$\,Gyr, which are mostly $\alpha$-rich and metal-poor stars. Age  of $\alpha$-rich and metal-poor star has slight chance to be given by a value of smaller than 8\,Gyr, which produce a small relative age error.   There are a fraction of old RGBs ($\tau > 10$\,Gyr) with  very large relative age errors, which is the consequence of that these stars have very large mass errors compared to other RGB stars as shown in Fig.\,\ref{mass_error}.     }

%\textbf{As shown in Fig.\,\ref{age_error}, stars with $M < 0.7\,M_{\odot}$ are stars with $\tau > 11$\,Gyr and small age \textbf{uncertainties}. Eleven of these twelve post-mass-transfer helium-burning stars are stars with $\tau > 12$\,Gyr. }

\begin{figure*}
\centering
\includegraphics[width=6.0in]{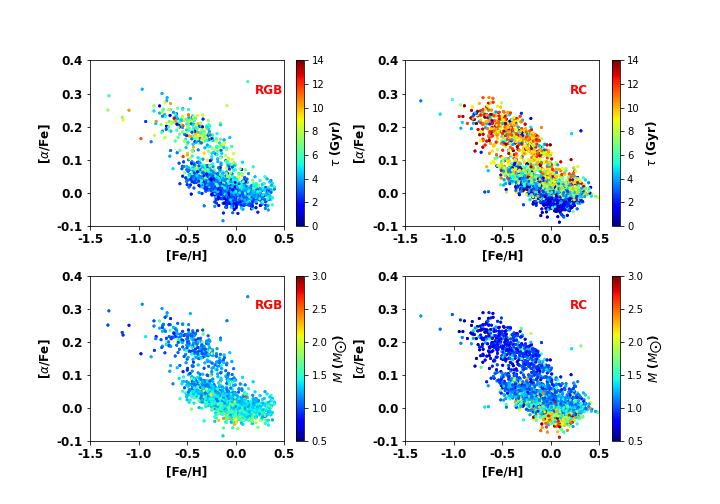}
\caption{Distributions of LAMOST-Kepler 1,493 training RGB stars (left panels) and  2,499 training RC stars (right panels)  on the [Fe/H]--[$\alpha$/Fe] plane, colour-coded by stellar age (top panels) and mass (bottom panels).  Here the values of [Fe/H] and [$\alpha$/Fe] come from the value-added catalogue of LAMOST DR8 \citep{Wang2022}. }
\label{train_age_mass}
\end{figure*}

After estimating the masses and ages of these 3,103 and  3,114 RGB and RC common stars, we divide both RGB and RC common stars into two sub-samples, a training and a testing sub-sample.  The training sample of RGB and RC  contains 1,493 and 2,499 stars,  respectively.   Other 1,610 and 615 RGB and RC stars form the testing samples.   When selecting training stars, the number of training stars in each age bin is the same as each other as much as possible. As shown in Fig.\,\ref{age_error}, the RGB sample contains fewer old stars as compared to the RC sample.  This is the same for the training set.   Fig.\,\ref{train_age_mass} shows the distributions of the RGB, RC training samples and testing samples on the [M/H]–[$\alpha$/Fe] plane, colour-coded by mass and age.  

\subsection{Age and mass determinations of RGB and RC stars in LAMOST DR8}
After building up the training and testing stars with accurate masses and ages, we derive masses and ages of RGB and RC stars from the LAMOST DR8 spectra, using the neural network method, which is detailed described in Section 3.1. 
The aforementioned RGB and RC training stars are adopted as training samples.  After pre-processing the LAMOST spectra (same as that of \citet{Wang2022}) of all the training and testing stars, we build up the model to construct the relation between the pre-processed LAMOST spectra and mass/age for RGBs and RCs.  Our neural network models to estimate masses and ages also  contain three layers, the  total number of neurons for
the first, second, and third layers are respectively 512, 256, and 64, which is the same as the neural network models to estimate asteroseismic parameters in Section 3.

\begin{figure*}
\centering
\includegraphics[width=6.0in]{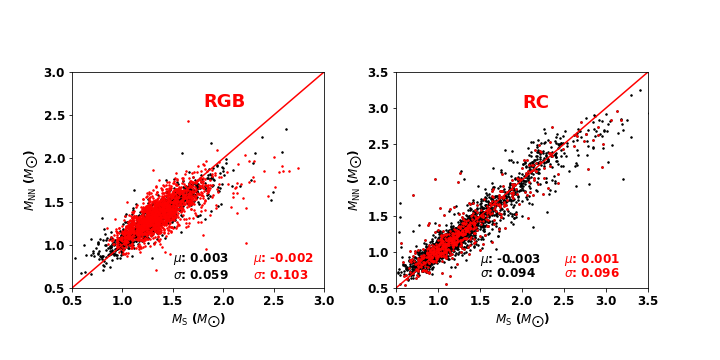}
\caption{The comparisons of mass estimated with scaling relation ($M_{\rm S}$)  in Section 4.1   and mass derived by neural network model ($M_{\rm NN}$) using LAMOST spectra for LAMOST-Kepler  RGB (left panel) and  RC (right panel) stars testing sample (red dots; 1,610 RGBs, 615 RCs) and training sample (black dots; 1,493 RGBs, 2,499 RCs). The $\mu$ and $\sigma$ are the mean and standard deviations of $(M_{\rm S}-M_{\rm NN})$/$M_{\rm S}$, as marked in the bottom-right corner of each panel. }
\label{mass_test}
\end{figure*}

\begin{figure*}
\centering
\includegraphics[width=6.0in]{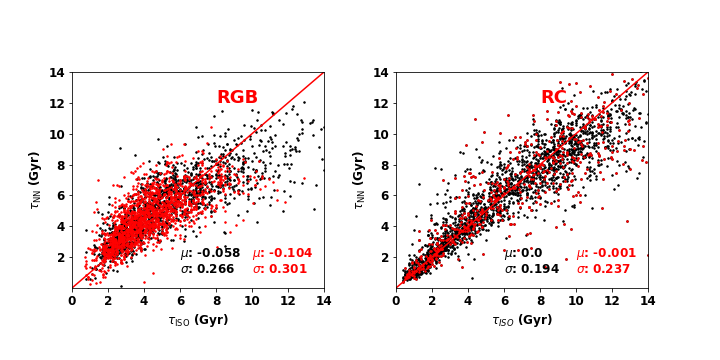}
\caption{The comparisons of age estimated with isochrone fitting ($\tau_{\rm ISO}$)  in Section 4.1   and age derived by neural network model ($\tau_{\rm NN}$) using LAMOST spectra for LAMOST-Kepler 3,103 RGB (left panel) and RC (right panel) stars in the testing sample (red dots; 1,610 RGBs, 615 RCs) and training sample (black dots; 1,493 RGBs, 2,499 RCs). The $\mu$ and $\sigma$ are the mean and standard deviations of $(\tau_{\rm ISO}-\tau_{\rm NN})$/$\tau_{\rm ISO}$,  as marked in the bottom-right corner of each panel. }
\label{age_test}
\end{figure*}

Fig.\,\ref{mass_test} shows the comparison of mass estimated with scaling relation ($M_{S}$)  in Section 4.1   and mass derived by neural network model ($M_{\rm NN}$) using LAMOST spectra for testing RGB and RC stars.  We can find that  $M_{\rm NN}$ match well with  $M_{S}$ for individual stars.  The systematic uncertainties are very small, the standard deviations of estimated mass are 10 per\,cent and 9.6 per\,cent for RGBs and RCs, respectively.   Fig.\,\ref{age_test} shows the comparison of age estimated with the isochrone fitting method ($\tau_{\rm ISO}$)  in Section 4.1   and age derived by neural network model ($\tau_{\rm NN}$) using LAMOST spectra for testing RGB and RC stars.  We can find that the $\tau_{\rm NN}$ is consistent with the  $\tau_{\rm ISO}$ for individual stars. The $\tau_{\rm NN}$ are overestimated about 10 per\,cent for RGB stars. The systematic uncertainties of  $\tau_{NN}$ of RCs are very small.  The standard deviations of estimated age are 30 per\,cent and 24 per\,cent for RGBs and RCs, respectively.  Figs.\,\ref{mass_test} or \ref{age_test} also show the comparisons between $M_{\rm NN}$ and $M_{S}$ or $\tau_{\rm NN}$ and $\tau_{\rm ISO}$ for training stars.  The $M_{\rm NN}$ ($\tau_{\rm NN}$) is well in agreement with $M_{S}$ ($\tau_{\rm ISO}$).  The standard deviations of estimated mass are respectively 5.9 per\,cent and 9.4 per\,cent for RGB and RC training stars. For age estimates of RGB and RC training stars, the standard deviations are 26.6 per\,cent and 19.4 per\,cent, respectively. 
In summary, our neural networks could derive precise mass and age for RGB and RC stars from LAMOST spectra.

Finally, we adopt our neural network models to  937,082, 244,458, and 167,105 RGB, primary RC and secondary RC stellar spectra and derive their spectroscopic masses and ages.  

%\section{Ages and masses of MSTOs}
%For all selected MSTOs, we estimate their ages using a Bayesian approach similar with \cite{Xiang2017}  and \cite{Huang2020}.  The input stellar parameters contains the effective temperature $T_{\rm eff}$, metallicity [M/H] and Gaia EDR3 $G$ band absolute magnitudes estimated from the LAMOST spectra, provided by Wang et al. 2021.  As discussed in Section\,3.1, we adopt the PARSEC isochrones calculated with a mass-loss parameter $\eta Reimers = 0.2$ \citep{Bressan2012} as the stellar isochrones.

%In order to show the age estimations, posterior probability distributions as a function of age for three stars of typical ages for MSTO stars are  shown in Fig.\,\ref{age_pdf_msto}.  As shown in Fig.\,\ref{age_pdf_msto}, we can find that the distributions show prominent peaks, which suggest that the resultant ages are well constrained. 
%  Fig.\,\ref{} show the  distributions of ages of MSTOs  estimated using isochrone fitting method, and of the associated errors.  We can find that the typical age uncertainties of  MSTO stars are ... per\,cent.  There are only a small fraction  stars have age uncertainties larger than 40 per\,cent. 

%\begin{figure*}
%\centering
%\includegraphics[width=5.5in]{figure/age_pdf_msto.pdf}
%\caption{The posterior probability distributions as a function of age for three stars of typical ages for MSTO stars. The blue solid lines denote the medians values of the stellar ages.  The blue dashed lines show the standard deviations of  the stellar ages. }
%\label{age_pdf_msto}
%\end{figure*}

\section{Validation of the spectroscopic masses and ages of RGBs and  RCs}
In this section, we first examine the uncertainties of the masses and ages of RGBs and RCs using the neural network method through internal comparisons. Secondly, through external comparisons, we further examine the uncertainties, including the systematic biases, of masses and ages.

\subsection{The uncertainties of estimated masses and ages of RGBs and RCs}
 The uncertainties of estimated masses and ages depend on the spectral noise (random uncertainties) and method uncertainties. 

\begin{figure*}
\centering
\includegraphics[width=6.0in]{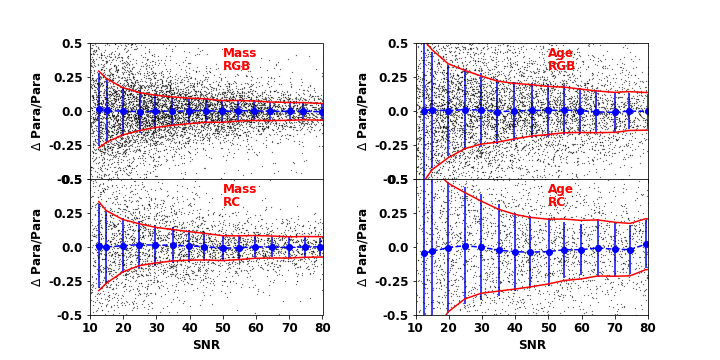}
\caption{Relative internal residuals of estimated masses (left panels) and ages (right panels)  using neural network method given by duplicate observations of similar spectral SNRs for LAMOST RGB (top panels) and RC (bottom panels) stars.  Black dots are the differences of duplicate observations of SNR
differences smaller than 10\,\%. Blue dots and error bars represent the median values and standard deviations (after divided by  $\sqrt{2}$) of the relative residuals in the individual spectral SNR bin. Red lines indicate fits of the standard deviations as a function of spectral SNRs. }
\label{para_snr}
\end{figure*}

As discussed in \cite{Huang2020} and \cite{Wang2022}, the random uncertainties of masses and ages are estimated by comparing results derived from duplicate observations of similar spectral SNRs (differed by less than 10\,per\,cent) collected during different nights.  The relative residuals of mass and age estimate  (after
divided by $\sqrt{2}$) along with mean spectral SNRs are shown in Fig.\,\ref{para_snr}.  To properly obtain  random uncertainties of age and mass, we fit the relative residuals with the  equation similar to \cite{Huang2020}:
\begin{equation}
\sigma_{r}=a+\frac{c}{(\rm SNR)^{\emph b}},
\end{equation}
where $\sigma_{r}$ represents the random uncertainty.   Besides the random uncertainties, method uncertainties ($\sigma_{m}$) are
also considered when we estimate the uncertainties of stellar parameters.  The final uncertainties are given by $\sqrt{\sigma_{r}^{2}+\sigma_{m}^{2}}$.  The method uncertainties are provided by the relative residuals between our estimated results and the asteroseismic results of the training sample.  For the age and mass of RGB stars, the method uncertainties are 26.6 and 5.9 per\,cent as shown in Figs.\,\ref{mass_test} and \ref{age_test}, respectively.  The method uncertainties of age and mass are 19.4 and 9.4 per\,cent for RC stars as shown in Figs.\,\ref{mass_test} and \ref{age_test}, respectively.  
%The resulting fit coefficients of random errors and method errors for different stellar parameters  are present in Table\,\ref{table_snrfit}.  

%\begin{table}
%\centering
%\caption{Fit coefficients of random errors and method errors of the mass and ages  uncertainty estimates of RGB and RC stars.}
%\begin{tabular}{ccccc}
%\hline
 %Parameters&a&b&c &$\sigma_{m}$\\
%\hline
%Mass\_RGB&0.039&1.302&6.612&0.059\\
%\hline
%Age\_RGB&0.080&1.094&6.981&0.266\\
%\hline
%Mass\_RC&0.058&1.536&13.221&0.094\\
%\hline
%Age\_RC&0.115&1.156&11.317&0.194\\
%\hline
%\end{tabular}
%\label{table_snrfit}
%\end{table}

%\subsubsection{The uncertainties of estimated masses and ages MSTOs}
%When we estimate the masses and ages of MSTOs through isochrone fitting method, the relative errors are also given. 

\subsection{The ages of members of open clusters}
\begin{figure*}
\centering
\includegraphics[width=6.0in]{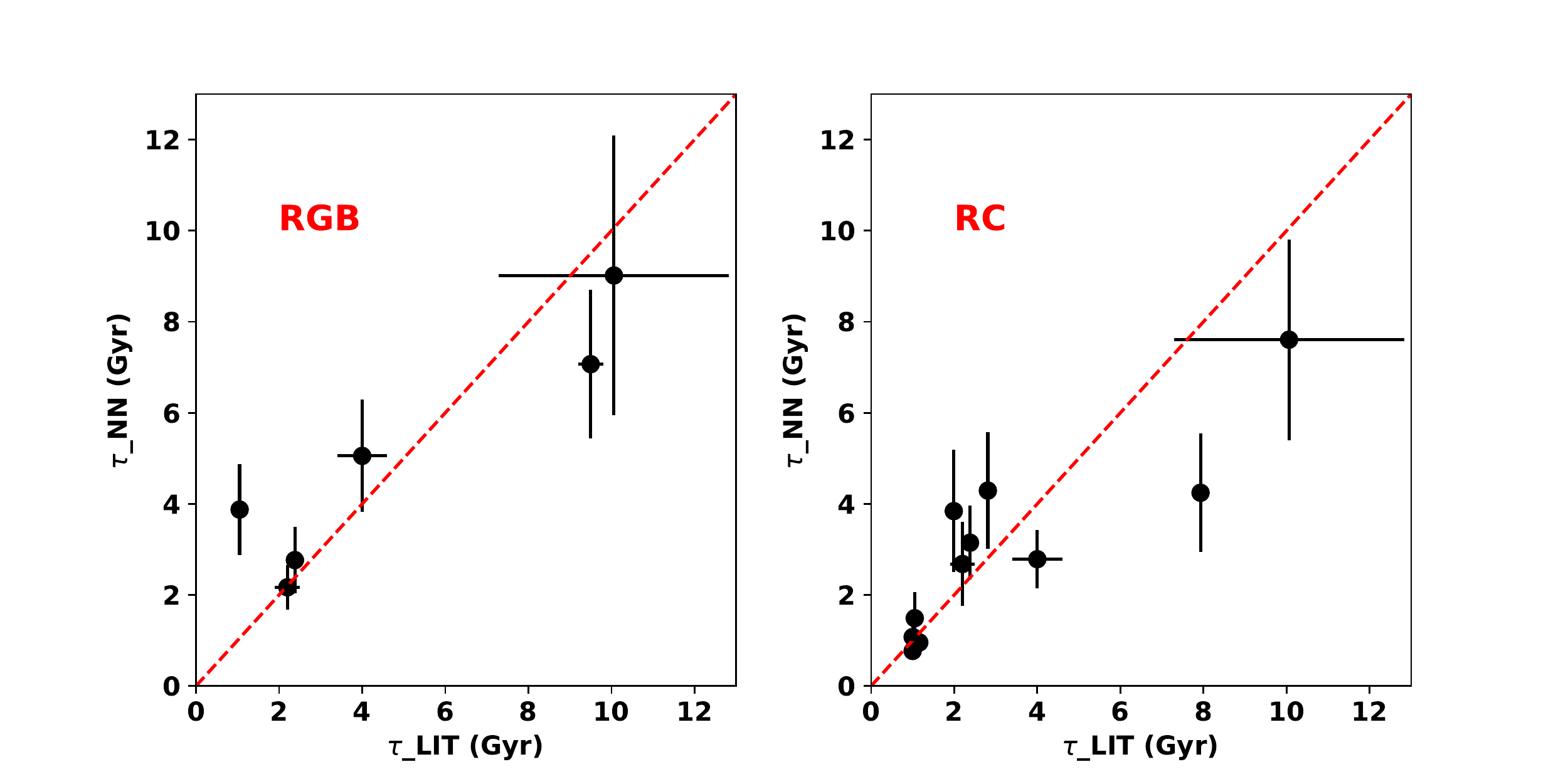}
\caption{Comparisons of our estimated ages ($\tau_{\rm NN}$) of open clusters with literature values ($\tau_{\rm LIT}$) for LAMOST RGBs (left panel) and RCs (right panel).}
\label{age_oc}
\end{figure*}

Stars in the open cluster are believed to form almost simultaneously from a single gas cloud, thus they have almost the same age, metallicity, distance and kinematics. Open cluster is a good test bed to check the accuracy of ages.  \cite{Zhongjing2020} and \cite{Zhongjingcatalog} provide 8,811 cluster members of 295 clusters observed by LAMOST, who cross-match the open cluster catalogue and the member stars provided by \cite{Cantat-Gaudin2018} with LAMOST\,DR5.  Amongst these clusters, open clusters with a wide age range (1--10\,Gyr) are carefully selected for checking our age estimates.   As shown in Fig.\,\ref{age_oc}, we can find that our ages of open clusters are consistent with these of literature for both RGB and RC stars from the young part ($\sim$ 2\,Gyr) to the old part ($\sim 10$\,Gyr).  Table.\,\ref{table_oc_age} present our age estimates and literature age values.

\begin{table*}[!ht]
	\centering
	\caption{Age values of open clusters estimated using our neural network and ages of literature values for both LAMOST  RGB and RC stars. The numbers of RGB and RC cluster members are also presented. }
	\begin{threeparttable}
\begin{tabular}{ccccccc}
	\hline
 	Cluster&LIT age (Gyr)&age of RGB (Gyr)&number of RGB& age of RC (Gyr)& number of RC  & References \tnote{*} \\
	\hline
	NGC\,6811&1.00$\pm$0.10&--&0&0.76$\pm$0.19&6&1, 2\\
	\hline
	NGC\,2420&2.20$\pm$0.30&2.16$\pm$0.48&1&2.67$\pm$0.92&4 & 3\\
	\hline
	NGC\,6819&2.38$\pm$0.22&2.76$\pm$0.73&6&3.14$\pm$0.81&14 & 4, 5\\
	\hline
	NGC\,2682&4.00$\pm$0.60&5.05$\pm$1.23&6&2.78$\pm$0.64&2 & 6\\
	\hline
	NGC\,6791&9.50$\pm$0.30&7.06$\pm$1.61&2&--&--& 7\\
	\hline
	Berkeley\,17&10.06$\pm$2.77&9.01$\pm$3.07&2&7.6$\pm$2.04&2 & 3, 8\\
	\hline
	Berkeley\,21&1.99&--&0&3.84$\pm$1.34&2 & 9--11\\
	\hline
	Berkeley\,39&7.93&--&0&4.24$\pm$1.29&2 & 9--11\\
	\hline
	Berkeley\,78&2.81&--&0&4.29$\pm$1.28&2 & 9--11\\
	\hline
	Czernik\,27&1.16&--&0&0.96$\pm$0.26&2 & 9--11\\
	\hline
	FSR\,0850&1.0$\pm$0.05&--&0&1.07$\pm$0.38&2 & 9--11\\
	\hline
	NGC\,1245&1.05&3.87$\pm$1.00&5&1.49$\pm$0.56&12 & 9--11\\
	\hline
	
\end{tabular}
\end{threeparttable}
\begin{tablenotes}
\footnotesize
 
\item[*] \textbf{Notes. *} References: (1) \cite{Sandquist2016}; (2) \cite{Molenda2014}; (3) \cite{Salaris2004};  (4) \cite{Anthony2014}; (5) \cite{Brewer2016}; (6) \cite{Stello2016}; (7) \cite{An2015}; (8) \cite{Bragaglia2006}; (9) \cite{Zhongjingcatalog}; (10) \cite{Kharchenko2013}; (11) \cite{Bossini2019}.
\end{tablenotes}
\label{table_oc_age}
\end{table*}

\subsection{Comparison of masses and ages of RGB and RC stars with previous results}
\begin{figure*}
\centering
\includegraphics[width=6.0in]{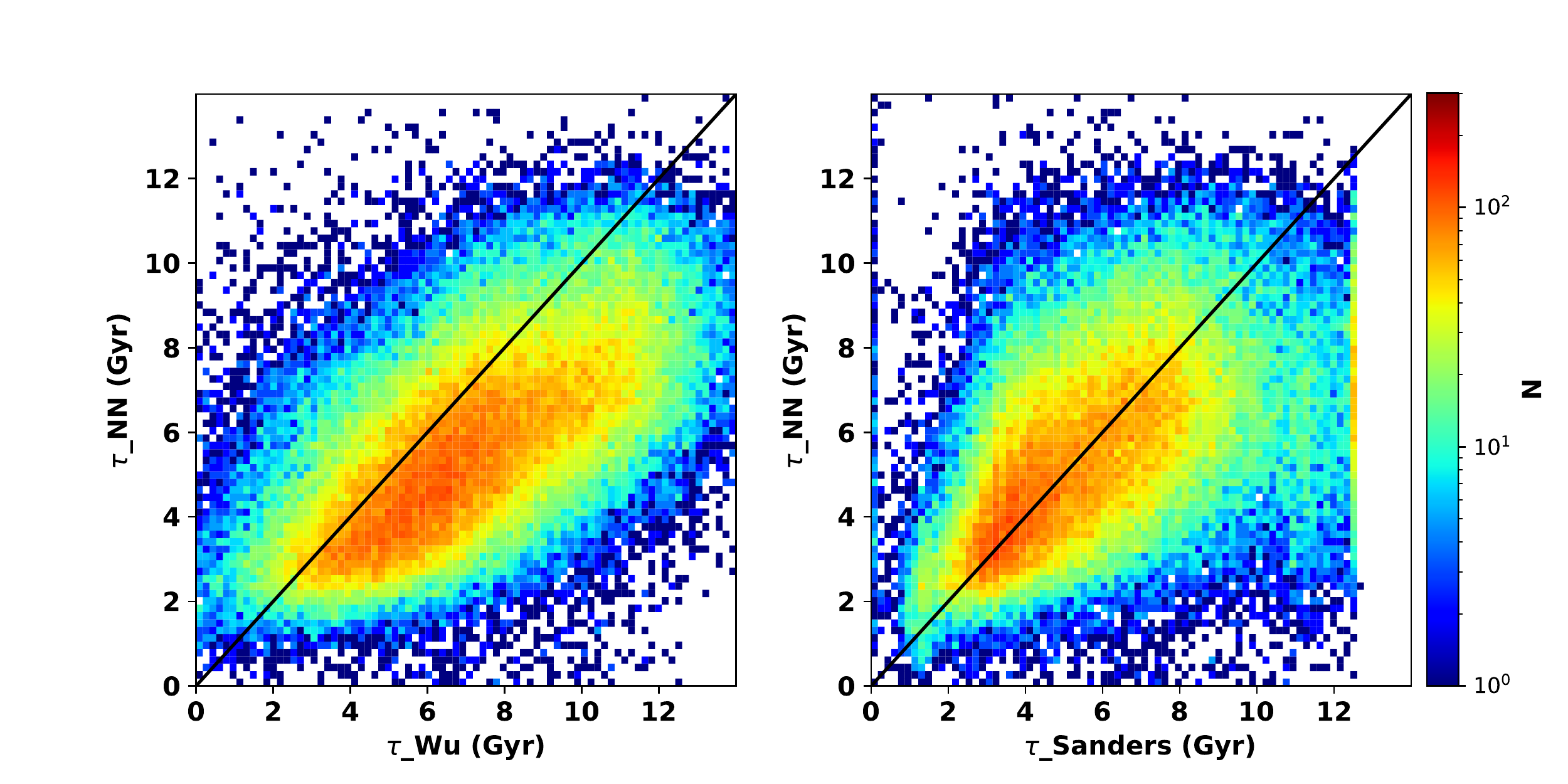}
\caption{Comparisons of our estimated ages with these provided by \citet{Wuyaqian2019} (left panel) and \citet{Sanders2018} (right panel) for 97,112 and 83,199 common RGB stars with $\rm SNRs > 50$, respectively.}
\label{age_wu_sanders}
\end{figure*}

%\cite{Wuyaqian2019} provide a catalogue of stellar ages and masses for $\sim$0.64 million RGB stars targeted by  the LAMOST Galactic Spectroscopic Survey (DR4). 
A catalogue containing accurate stellar ages for $\sim$0.64 million RGB stars targeted by the LAMOST DR4 has been provided by  \cite{Wuyaqian2019}. 
 \cite{Sanders2018} has determined distances, and ages for $\sim$3 million stars with astrometric information from Gaia\,DR2 and spectroscopic parameters from massive spectroscopic surveys: including the APOGEE, Gaia-ESO, GALAH, LAMOST, RAVE, and SEGUE surveys.
 We cross-match our catalogue with their catalogues and obtain $\sim$ 97,000 and 83,000  common RGB stars (SNR$\geq$50, 2.0 $<\log g< $ 3.2\,dex, 4000 $<T_{\rm eff}<$ 5500\,K) with the catalogue of \cite{Wuyaqian2019} and \cite{Sanders2018}, respectively.  
 The comparisons are shown in Fig.\,\ref{age_wu_sanders}. Generally,  our ages are younger than those estimated by \cite{Wuyaqian2019}.  In particular, for the old parts,  our ages are younger than the ages of   \cite{Wuyaqian2019} with 2--3 Gyr.  This is mainly the different choices of isochrone database between this work and \cite{Wuyaqian2019}, who adopt the isochrones from PARSEC and Yonsei–Yale  \citep[$\rm Y^{2}$;][]{Demarque2004}, respectively.  The lack of old (low mass: $M < 1.0\,M_{\odot}$) RGB training stars in this work may also cause our relatively younger age estimates.
 Our estimated ages could agree well with these estimated by \cite{Sanders2018}.  

\begin{figure*}
\centering
\includegraphics[width=6.0in]{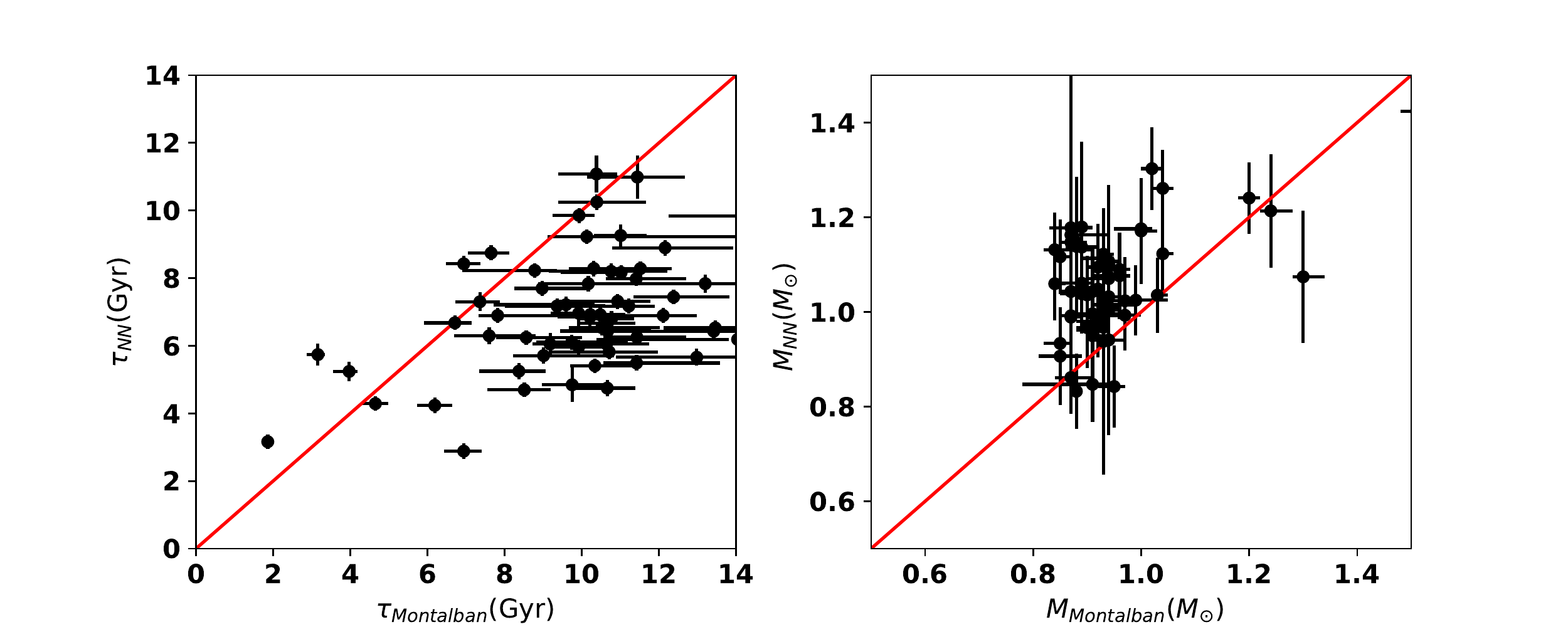}
\caption{Comparisons of our estimated ages and masses with these provided by \citet{Montalban2021} for 62 common  RGB stars.}
\label{age_montalban}
\end{figure*}

\cite{Montalban2021} has provided precise stellar ages ($\sim 11$\%) and masses for 95 RGBs observed by the Kepler space mission by combining the asteroseismology with kinematics and chemical abundances of stars.  There are 62 common stars between our catalogue and their list. As shown in Fig.\,\ref{age_montalban}, ages of old stars with $\tau > 8$\,Gyr are underestimated by 2--3\,Gyr compared to the results of \cite{Montalban2021}, because masses of these stars are overestimated by $\sim 0.1 M_{\odot}$.  \cite{Montalban2021} estimate the masses adopting the individual frequencies of radial modes as observational constraints, which may provide more accurate masses \citep{Lebreton2014, Miglio2017, Rendle2019,Montalban2021}. Building up a training sample containing enough RGBs with accurate age and mass estimates adopting individual frequencies as observational constraints may dramatically improve the determination of age and mass for RGBs using machine learning methods.  A potential challenge is the lack of precise individual frequencies for a larger sample size of red giant stars ($\sim$2000--6000 stars).   Yet this is out of the topic in this paper.
%Previous works \citep{Lebreton2014, Miglio2017, Rendle2019,Montalban2021} have argued that the mean properties of stellar oscillations (such as the large frequency separation) are not enough to derive accurate masses and ages, individual frequencies are needed to reduce uncertainties affecting estimated masses and ages with respect to the precision and accuracy resulting from scaling relations. 

\begin{figure*}
\centering
\includegraphics[width=6.0in]{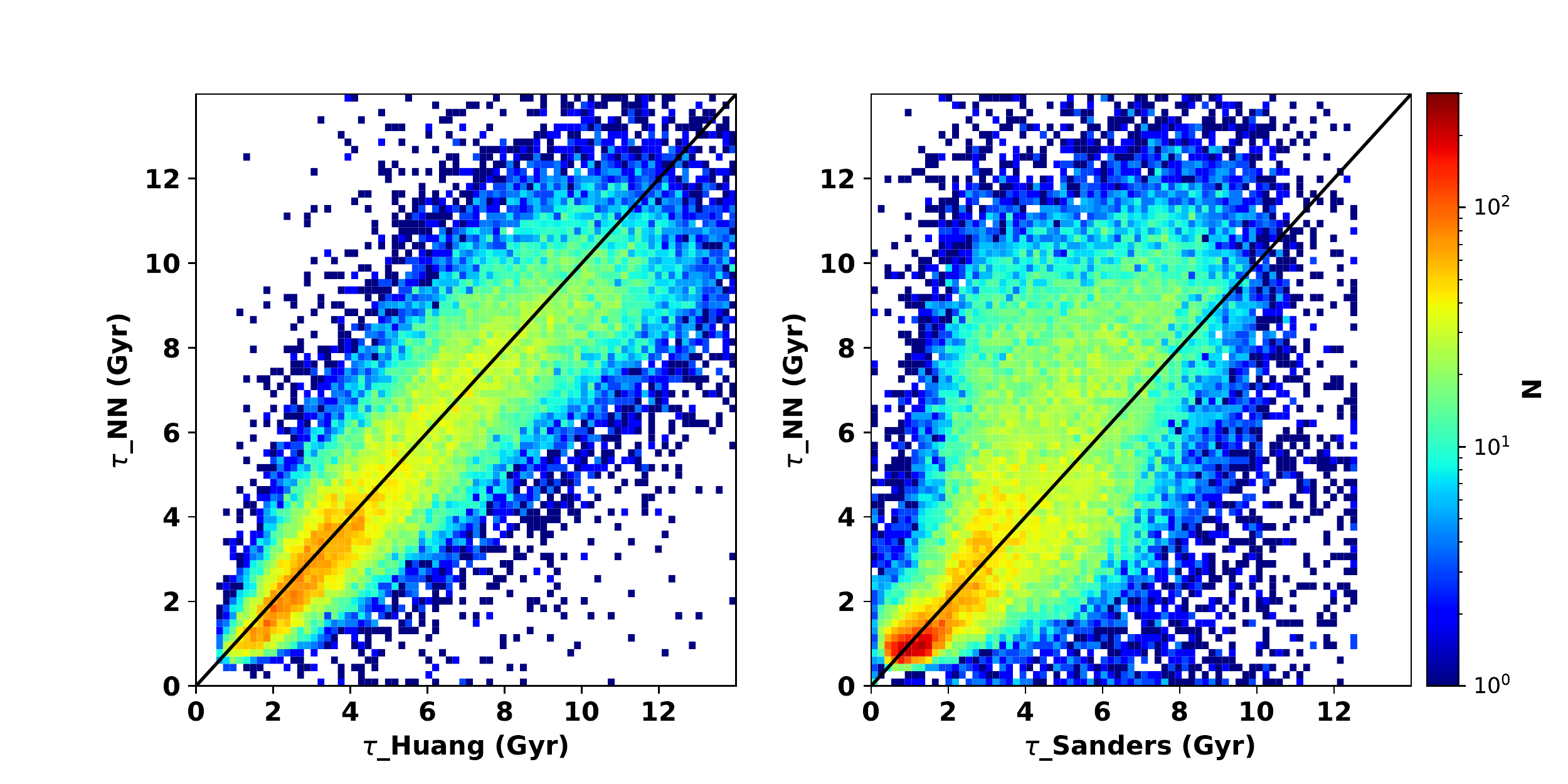}
\caption{Comparisons of our estimated ages with these provided by \citet{Huang2020} (left panel) and \citet{Sanders2018} (right panel) for 42,969 and 52,944 common RC stars with $\rm SNRs > 50$, respectively.}
\label{age_huang_sanders}
\end{figure*}

%\subsection{Comparison of ages of RC stars with previous results}
The precise distances, masses, ages and 3D velocities for $\sim$140,000 primary RCs selected from the LAMOST have been determined by our former effort \citep{Huang2020}. 
The catalogue of \cite{Sanders2018} also contain a lot of  RC stars with age estimates.    
%Through cross-matching our catalogue and their catalogues,  we obtain $\sim$ 52,000 and 72,000  RC stars (SNR$\geq$50) in common with the catalogue of \cite{Huang2020} and \cite{Sanders2018}, respectively.  
Our catalogue has $\sim$ 52,000 and 72,000  RC stars (SNR$\geq$50) in common with the catalogue of \cite{Huang2020} and \cite{Sanders2018}.
Fig.\,\ref{age_huang_sanders} shows the comparisons between ages in this work and the ages of   \cite{Huang2020} and \cite{Sanders2018} for RC stars.  Our ages are consistent very well with these of  \cite{Huang2020}.  The ages in this work are generally consistent with these of \cite{Sanders2018}, although a mild offset of 2--3\,Gyr (this work minus \cite{Sanders2018}) is detected.

In summary,  the uncertainties of ages and masses of RC stars are smaller than 25 and 10 per\,cent for stars with spectral SNR $> 50$ through the internal comparisons, respectively.  For RGB stars with spectral SNR $ > 50$, the uncertainties of ages and masses are respectively smaller than 20 and  8 per\,cent.  The systematic uncertainties of our age estimates are very small for both RGB and RC stars,  by comparing the age derived in this work to the literatures for open cluster member stars. 
%The uncertainties of ages and masses of RGB stars are smaller 20 and  8 per\,cent for stars with SNR $> 50$ through the internal \textbf{comparisons}, respectively.  
Our age estimations match well with these of \citet{Sanders2018} for RGB stars and these of \citet{Huang2020} for RC stars. 

\section{Properties of the LAMOST RGB and RC samples}
Besides the accurate masses and ages estimated here, the stellar atmospheric parameters ($T_{\rm eff}$, $\log\,g$ and [Fe/H]), chemical element abundance ratios ([M/H], [$\alpha$/M], [C/Fe] and [N/Fe]) and 14 bands absolute magnitudes ($M_{G}, M_{Bp}, M_{Rp}$ of Gaia bands, $M_{J}, M_{H}, M_{Ks}$ of 2MASS bands, $M_{W1}, M_{W2}$ of WISE bands, $M_{B}, M_{V}, M_{r_{A}}$ of APASS bands and $M_{g}, M_{r}, M_{i}$ of SDSS bands) and photometric distances are also provided, which are taken from the value-added catalogue of LAMOST\,DR8 built up by \cite{Wang2022}.
Thus, we could study the spatial coverage, and age distributions on the $R$--$Z$ plane and on the [Fe/H]--[$\alpha$/Fe] plane of our RGB, primary RC and secondary RC samples. 

\begin{figure*}
\centering
\includegraphics[width=6.0in]{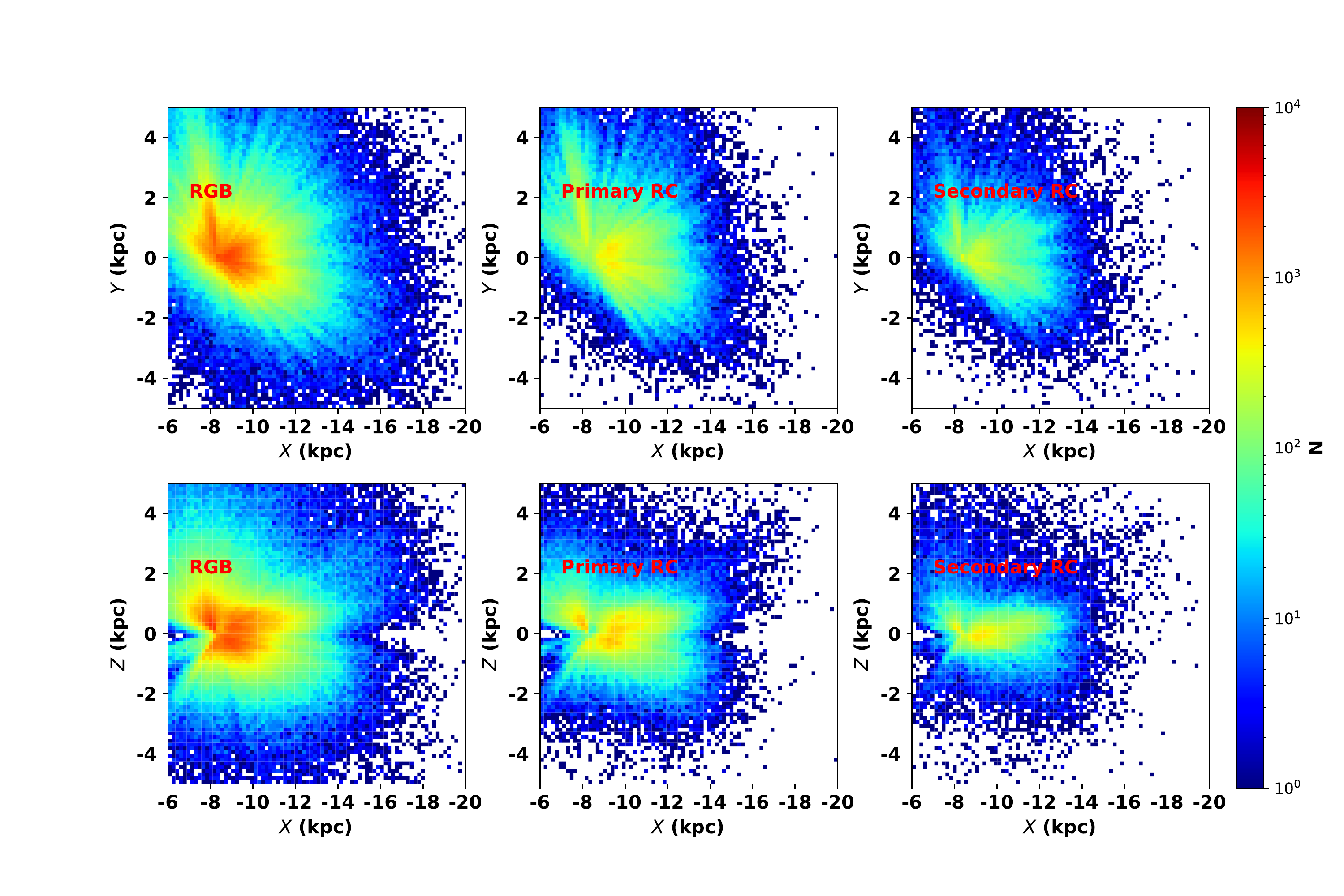}
\caption{Stellar number density distributions of the 520,073 unique LAMOST RGB (left panels), 154,587 unique primary RC (middle panels) and 66,395 unique secondary RC (right panels) stars with spectral $\rm SNRs> 20$ on the  $X$--$Y$ (top panels) and $X$--$Z$ (bottom panels) planes in Galactic Cartesian coordinates. }
\label{xyxz}
\end{figure*}

%\subsection{The spatial coverage of the LAMOST \textbf{RGB, primary RC and secondary RC} samples}
Fig.\,\ref{xyxz} shows the stellar number density distributions of our RGB, primary RC and secondary RC stellar samples on the $X$--$Y$ and $X$--$Z$ planes, which cover a large volume of the Milky Way ($5 < R < 20$\,kpc and $|Z| <5$\,kpc). 
One can use it to probe the structural, chemical, and kinematic properties of both the Galactic thin, thick discs and halo combining the proper motions provided by Gaia DR3 \citep{gaiadr3, gaiadr3cata}.  

\begin{figure*}
\centering
\includegraphics[width=6.0in]{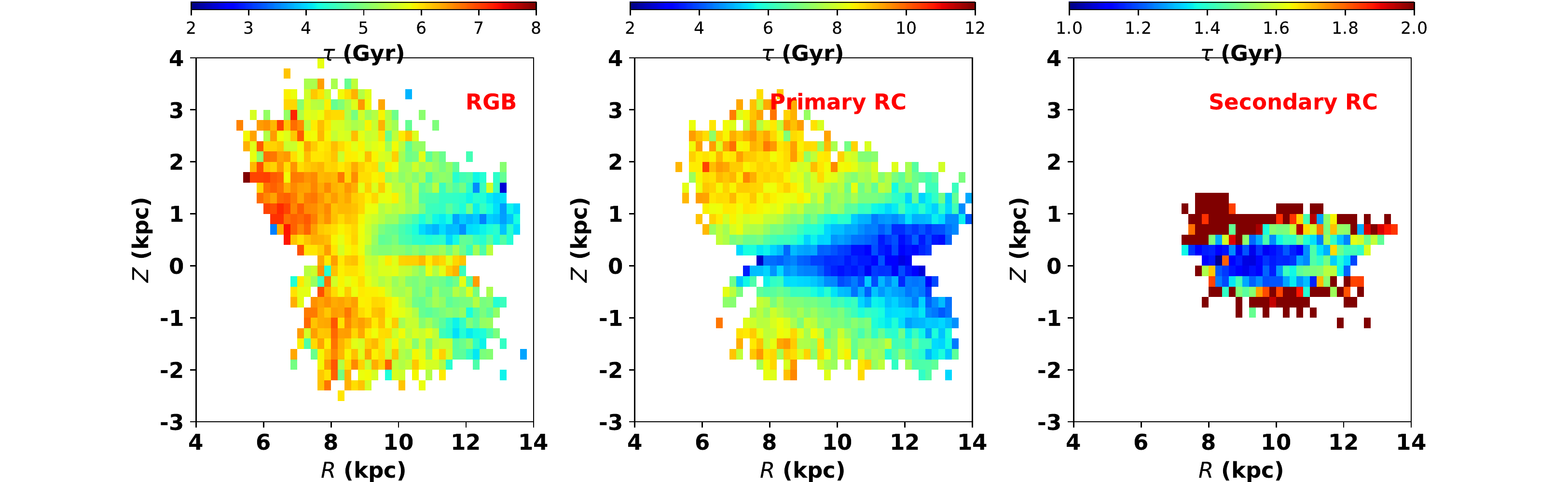}
\caption{The median age distributions of the 259,372 unique LAMOST RGB (left panels), 92,795 unique primary RC (middle panels) and 2,8104 unique secondary RC (right panels) stars with spectral $\rm SNRs> 50$ on the  $R$--$Z$ plane. }
\label{age_distribution}
\end{figure*}

%\subsection{The age distributions  of the LAMOST RGB, \textbf{primary RC and secondary RC} samples on the $R$--$Z$ plane}

The age distributions of RGB, primary RC and secondary RC stars on the $R$--$Z$ plane are presented in Fig.\,\ref{age_distribution}.  From this figure, we can see that there are negative age gradients in the radial direction and positive age gradients in the vertical direction for both RGB and primary RC stars, which are reported in a series of previous studies \citep[e.g.,][]{Martig2016, Casagrande2016}.  For secondary RCs, there are positive age gradients in the vertical direction. Flat or slightly positive radial age gradients are found for secondary RCs, which may be meaningless considering the small mean age coverage (from $\sim$ 1.1\,Gyr at $R \sim 9$\,kpc to 1.5\,Gyr at $R \sim $12\,kpc) and large age uncertainties at youngest part.    The strong flaring of young Galactic disc (young stellar populations extended to higher heights from the Galactic plane with $R$ increasing) is clearly seen in the age distributions of RGB and primary RC stars.  Similar results are also noted by  \cite{Xiang2017}, \cite{Wuyaqian2019} and \cite{Huang2020}.

%\subsection{Stellar number density and age distributions of the LAMOST RGB, \textbf{primary RC and secondary RC} samples on the [Fe/H]-[$\alpha$/Fe]}

\begin{figure*}
\centering
\includegraphics[width=6.0in]{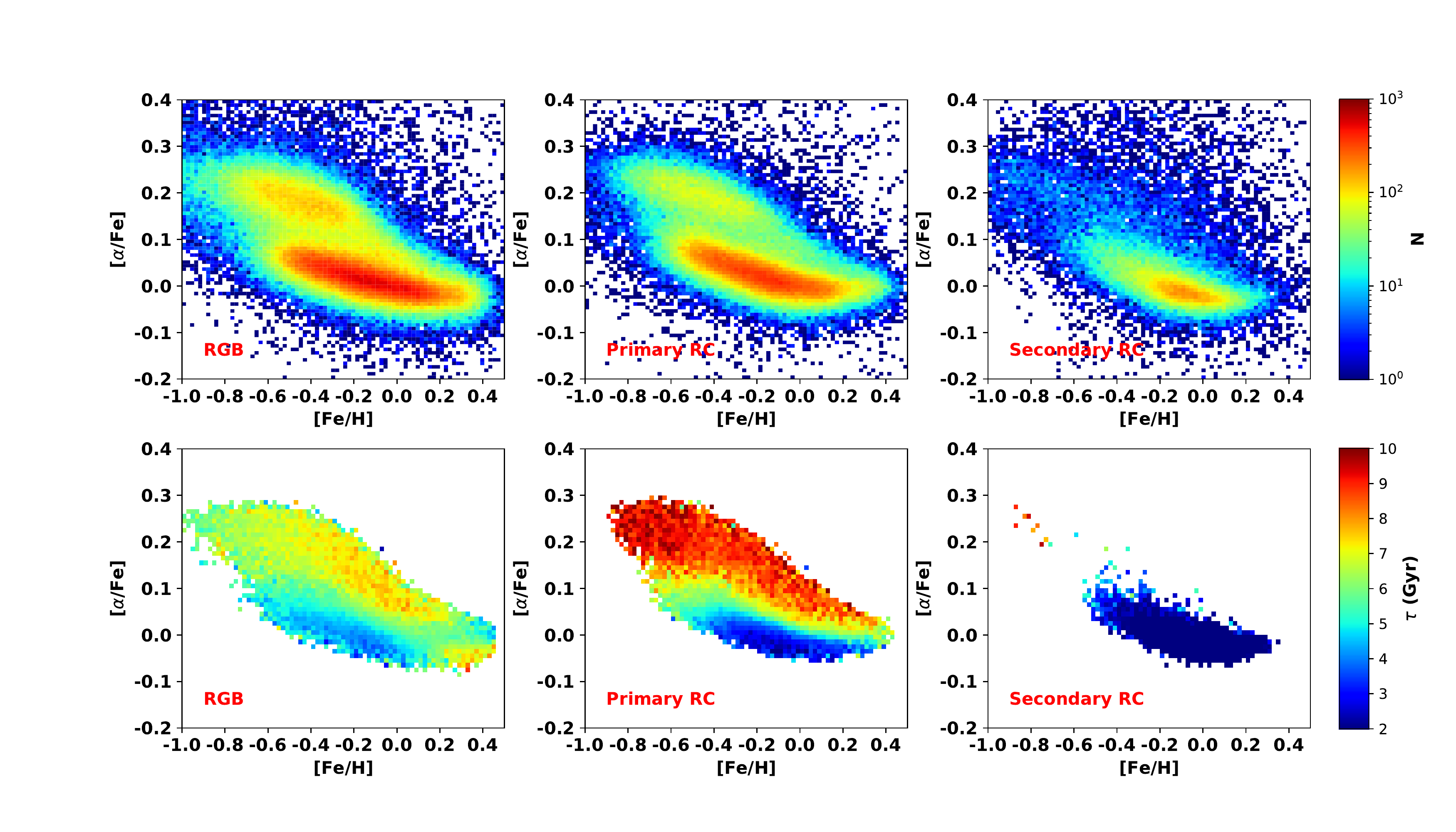}
\caption{The stellar number density (top panels) and mean age (bottom panels) distributions on the [Fe/H]--[$\alpha$/Fe] plane for the 259,372 unique LAMOST RGB (left panels), 92,795 unique primary RC (middle panels) and 2,8104 unique secondary RC (right panels) stars with spectral $\rm SNRs> 50$. }
\label{age_abundance}
\end{figure*}

Fig.\ref{age_abundance} shows the stellar number density and mean age distributions on the [Fe/H]--[$\alpha$/Fe] plane for the RGB, primary RC and secondary RC stars.  The stellar number density distributions of RGB and primary RC stars suggest that the RGB and primary RC stellar populations could be clearly distinguished into two individual sequences, the high-$\alpha$  and low-$\alpha$ sequences.  The high- and low-$\alpha$ sequences are associated with the so-called chemically thick and thin discs.  For primary RC stellar populations, the mean age of high- and low-$\alpha$ sequence are $\sim$\,8\,Gyr and $\sim 2.5$\,Gyr, respectively.  The high-$\alpha$ sequence is older than the low-$\alpha$ sequence.  The results are consistent with previous findings for the solar neighbourhood using high-resolution spectroscopy \citep{Bensby2003,Haywood2008,Haywood2013,Hayden2015} and results using a large sample of stars with ages derived from the LAMOST low-resolution spectra \citep{Xiang2017, Wuyaqian2019, Huang2020}.  For RGB stellar populations, the age of high- and low-$\alpha$ sequences are $\sim 6$\,Gyr and $\sim 3$ Gyr, respectively. The high-$\alpha$ sequence is older than the low-$\alpha$ sequence with a smaller age difference as compared to that of primary RC stellar populations.  For RGB stellar populations, the eldest stellar populations are located in the transition region of high- and low-$\alpha$ sequences. 
Almost all of secondary RC stars are thin disc stars, because they are massive young stars. 
 %Stars with metallicity higher than 0.2\,dex and age older than that of low-$\alpha$ sequence are also found in our RGB stellar population.  

For both RGB and primary RC stellar populations,  there are some stars with metallicity higher than 0.2\,dex and older than  low-$\alpha$ sequence. 
Some of these stars are the so-called ``old yet metal-rich" stars ($6 < \tau < 8$\,Gyr, $[\rm Fe/H] > 0.2$\,dex, $-0.05 < [\alpha/\rm Fe] < 0.05$\,dex).  It has been suggested that these   ``old yet metal-rich" stars may come from the inner disc through radial migration processes \citep{Xiangmdf, Wang2019, Feuillet2018, Feuillet2019, Hasselquist2019, Boeche, Chenyq} . Old and metal-rich stars in the inner disc will move outward through the radial migration process and produce these ``old yet metal-rich" stars. 

\section{Summary}
In the current work,  we build up a catalogue of masses and ages for 937,082, 244,458, and 167,105 RGBs, primary RCs and secondary RCs selected from LAMOST\,DR8 low-resolution stellar spectra of spectral SNR\,$>$\,10.  Amongst them, 696,680 RGBs, 180,436 primary RCs, and 120,907 secondary RCs are unique stars.   The RGBs, primary RCs and secondary RCs are identified with the large frequency spacing ($\Delta \nu$) and period spacing ($\Delta P$) estimated using a neural network method. The contaminations of RGB stars from RC stars and contaminations of RCs from RGBs are both about 1 per\,cent for stars with spectral SNR\,$>$\,20. 

Using the relations of stellar spectra and the masses and ages built up by neural network method using the RGB and RC stars in the LAMOST-Kepler fields that have accurate asteroseismic mass and age measurements as training samples, the masses and ages of all LAMOST RGB and RC stars are estimated.   The typical uncertainties of estimated masses and ages, as examined by the testing sample, are respectively 10\% and 30\% for RGB stars, and respectively 9\% and 24\% for RC stars.  

The RGB and RC stellar populations cover a large volume of the Milky Way ($5 < R < 20$\,kpc and $|Z| <5$\,kpc).   The stellar atmospheric parameters, chemical element abundances, 14 bands absolute magnitudes and photometric distances are also given here. One can use it to probe the structural, chemical, and kinematic properties of both the Galactic thin, thick discs and halo combining the proper motions provided by Gaia DR3.  The catalogue is available at the CDS and the LAMOST official website \footnote{The catalogue will be available at CDS via https://cdsarc.cds.unistra.fr/cgi-bin/qcat?J/A+A/ and the LAMOST official website via http://www.lamost.org/dr8/v1.0/doc/vac.}.
\begin{acknowledgements}
This work was funded by the National Key R\&D Program of China (No. 2019YFA0405500) and the National Natural Science Foundation of China (NSFC Grant No.12203037, No.12173007 and No.11973001).  We acknowledge the science research grants from the China Manned Space Project with NO. CMS-CSST-2021-B03.  We used data from the European Space Agency mission Gaia (http://www.cosmos.esa.int/gaia), processed by the Gaia Data Processing and Analysis Consortium (DPAC; see http://www.cosmos.esa.int/web/gaia/dpac/consortium). Guoshoujing Telescope (the Large Sky Area Multi-Object Fiber Spectroscopic Telescope LAMOST) is a National Major Scientific Project built by the Chinese Academy of Sciences. Funding for the project has been provided by the National Development and Reform Commission. LAMOST is operated and managed by the National Astronomical Observatories, Chinese Academy of Sciences.
\end{acknowledgements}

\bibliographystyle{aa}
\bibliography{age_mass}

\begin{thebibliography}{82}
\expandafter\ifx\csname natexlab\endcsname\relax\def\natexlab#1{#1}\fi

\bibitem[{{An} {et~al.}(2015){An}, {Terndrup}, {Pinsonneault}, \&
  {Lee}}]{An2015}
{An}, D., {Terndrup}, D.~M., {Pinsonneault}, M.~H., \& {Lee}, J.-W. 2015, \apj,
  811, 46

\bibitem[{{Anthony-Twarog} {et~al.}(2014){Anthony-Twarog}, {Deliyannis}, \&
  {Twarog}}]{Anthony2014}
{Anthony-Twarog}, B.~J., {Deliyannis}, C.~P., \& {Twarog}, B.~A. 2014, \aj,
  148, 51

\bibitem[{{Babusiaux} {et~al.}(2022){Babusiaux}, {Fabricius}, {Khanna},
  {Muraveva}, {Reyl{\'e}}, {Spoto}, {Vallenari}, {Luri}, {Arenou}, {Alvarez},
  {Anders}, {Antoja}, {Balbinot}, {Barache}, {Bauchet}, {Bossini}, {Busonero},
  {Cantat-Gaudin}, {Carrasco}, {Dafonte}, {Diakite}, {Figueras},
  {Garcia-Gutierrez}, {Garofalo}, {Helmi}, {Jimenez-Arranz}, {Jordi},
  {Kervella}, {Kostrzewa-Rutkowska}, {Leclerc}, {Licata}, {Manteiga}, {Masip},
  {Monguio}, {Ramos}, {Robichon}, {Robin}, {Romero-Gomez}, {Saez}, {Santovena},
  {Spina}, {Torralba Elipe}, \& {Weiler}}]{gaiadr3cata}
{Babusiaux}, C., {Fabricius}, C., {Khanna}, S., {et~al.} 2022, arXiv e-prints,
  arXiv:2206.05989

\bibitem[{{Bedding} {et~al.}(2011){Bedding}, {Mosser}, {Huber},
  {Montalb{\'a}n}, {Beck}, {Christensen-Dalsgaard}, {Elsworth}, {Garc{\'\i}a},
  {Miglio}, {Stello}, {White}, {De Ridder}, {Hekker}, {Aerts}, {Barban},
  {Belkacem}, {Broomhall}, {Brown}, {Buzasi}, {Carrier}, {Chaplin}, {di Mauro},
  {Dupret}, {Frandsen}, {Gilliland}, {Goupil}, {Jenkins}, {Kallinger},
  {Kawaler}, {Kjeldsen}, {Mathur}, {Noels}, {Silva Aguirre}, \&
  {Ventura}}]{Bedding2011}
{Bedding}, T.~R., {Mosser}, B., {Huber}, D., {et~al.} 2011, \nat, 471, 608

\bibitem[{{Bellinger}(2020)}]{Bellinger2020}
{Bellinger}, E.~P. 2020, \mnras, 492, L50

\bibitem[{{Bensby} {et~al.}(2003){Bensby}, {Feltzing}, \&
  {Lundstr{\"o}m}}]{Bensby2003}
{Bensby}, T., {Feltzing}, S., \& {Lundstr{\"o}m}, I. 2003, \aap, 410, 527

\bibitem[{{Boeche} {et~al.}(2013){Boeche}, {Siebert}, {Piffl}, {Just},
  {Steinmetz}, {Sharma}, {Kordopatis}, {Gilmore}, {Chiappini}, {Williams},
  {Grebel}, {Bland-Hawthorn}, {Gibson}, {Munari}, {Siviero}, {Bienaym{\'e}},
  {Navarro}, {Parker}, {Reid}, {Seabroke}, {Watson}, {Wyse}, \&
  {Zwitter}}]{Boeche}
{Boeche}, C., {Siebert}, A., {Piffl}, T., {et~al.} 2013, \aap, 559, A59

\bibitem[{{Bossini} {et~al.}(2019){Bossini}, {Vallenari}, {Bragaglia},
  {Cantat-Gaudin}, {Sordo}, {Balaguer-N{\'u}{\~n}ez}, {Jordi}, {Moitinho},
  {Soubiran}, {Casamiquela}, {Carrera}, \& {Heiter}}]{Bossini2019}
{Bossini}, D., {Vallenari}, A., {Bragaglia}, A., {et~al.} 2019, \aap, 623, A108

\bibitem[{{Bovy} {et~al.}(2014){Bovy}, {Nidever}, {Rix}, {Girardi}, {Zasowski},
  {Chojnowski}, {Holtzman}, {Epstein}, {Frinchaboy}, {Hayden}, {Rodrigues},
  {Majewski}, {Johnson}, {Pinsonneault}, {Stello}, {Allende Prieto}, {Andrews},
  {Basu}, {Beers}, {Bizyaev}, {Burton}, {Chaplin}, {Cunha}, {Elsworth},
  {Garc{\'\i}a}, {Garc{\'\i}a-Her{\'n}andez}, {Garc{\'\i}a P{\'e}rez},
  {Hearty}, {Hekker}, {Kallinger}, {Kinemuchi}, {Koesterke},
  {M{\'e}sz{\'a}ros}, {Mosser}, {O'Connell}, {Oravetz}, {Pan}, {Robin},
  {Schiavon}, {Schneider}, {Schultheis}, {Serenelli}, {Shetrone}, {Silva
  Aguirre}, {Simmons}, {Skrutskie}, {Smith}, {Stassun}, {Weinberg}, {Wilson},
  \& {Zamora}}]{Bovy2014}
{Bovy}, J., {Nidever}, D.~L., {Rix}, H.-W., {et~al.} 2014, \apj, 790, 127

\bibitem[{{Bragaglia} {et~al.}(2006){Bragaglia}, {Tosi}, {Andreuzzi}, \&
  {Marconi}}]{Bragaglia2006}
{Bragaglia}, A., {Tosi}, M., {Andreuzzi}, G., \& {Marconi}, G. 2006, \mnras,
  368, 1971

\bibitem[{{Bressan} {et~al.}(2012){Bressan}, {Marigo}, {Girardi}, {Salasnich},
  {Dal Cero}, {Rubele}, \& {Nanni}}]{Bressan2012}
{Bressan}, A., {Marigo}, P., {Girardi}, L., {et~al.} 2012, \mnras, 427, 127

\bibitem[{{Brewer} {et~al.}(2016){Brewer}, {Sandquist}, {Mathieu}, {Milliman},
  {Geller}, {Jeffries}, {Orosz}, {Brogaard}, {Platais}, {Bruntt}, {Grundahl},
  {Stello}, \& {Frandsen}}]{Brewer2016}
{Brewer}, L.~N., {Sandquist}, E.~L., {Mathieu}, R.~D., {et~al.} 2016, \aj, 151,
  66

\bibitem[{{Cantat-Gaudin} {et~al.}(2018){Cantat-Gaudin}, {Jordi}, {Vallenari},
  {Bragaglia}, {Balaguer-N{\'u}{\~n}ez}, {Soubiran}, {Bossini}, {Moitinho},
  {Castro-Ginard}, {Krone-Martins}, {Casamiquela}, {Sordo}, \&
  {Carrera}}]{Cantat-Gaudin2018}
{Cantat-Gaudin}, T., {Jordi}, C., {Vallenari}, A., {et~al.} 2018, \aap, 618,
  A93

\bibitem[{{Casagrande} {et~al.}(2016){Casagrande}, {Silva Aguirre},
  {Schlesinger}, {Stello}, {Huber}, {Serenelli}, {Sch{\"o}nrich}, {Cassisi},
  {Pietrinferni}, {Hodgkin}, {Milone}, {Feltzing}, \&
  {Asplund}}]{Casagrande2016}
{Casagrande}, L., {Silva Aguirre}, V., {Schlesinger}, K.~J., {et~al.} 2016,
  \mnras, 455, 987

\bibitem[{{Chaplin} \& {Miglio}(2013)}]{Chaplin2013}
{Chaplin}, W.~J. \& {Miglio}, A. 2013, \araa, 51, 353

\bibitem[{{Chen} {et~al.}(2019){Chen}, {Zhao}, {Zhao}, {Liang}, {Wu}, {Jia},
  {Tian}, \& {Liu}}]{Chenyq}
{Chen}, Y.~Q., {Zhao}, G., {Zhao}, J.~K., {et~al.} 2019, \aj, 158, 249

\bibitem[{{De Silva} {et~al.}(2015){De Silva}, {Freeman}, {Bland-Hawthorn},
  {Martell}, {de Boer}, {Asplund}, {Keller}, {Sharma}, {Zucker}, {Zwitter},
  {Anguiano}, {Bacigalupo}, {Bayliss}, {Beavis}, {Bergemann}, {Campbell},
  {Cannon}, {Carollo}, {Casagrande}, {Casey}, {Da Costa}, {D'Orazi}, {Dotter},
  {Duong}, {Heger}, {Ireland}, {Kafle}, {Kos}, {Lattanzio}, {Lewis}, {Lin},
  {Lind}, {Munari}, {Nataf}, {O'Toole}, {Parker}, {Reid}, {Schlesinger},
  {Sheinis}, {Simpson}, {Stello}, {Ting}, {Traven}, {Watson}, {Wittenmyer},
  {Yong}, \& {{\v{Z}}erjal}}]{DeSilva2015}
{De Silva}, G.~M., {Freeman}, K.~C., {Bland-Hawthorn}, J., {et~al.} 2015,
  \mnras, 449, 2604

\bibitem[{{Demarque} {et~al.}(2004){Demarque}, {Woo}, {Kim}, \&
  {Yi}}]{Demarque2004}
{Demarque}, P., {Woo}, J.-H., {Kim}, Y.-C., \& {Yi}, S.~K. 2004, \apjs, 155,
  667

\bibitem[{{Deng} {et~al.}(2012){Deng}, {Newberg}, {Liu}, {Carlin}, {Beers},
  {Chen}, {Chen}, {Christlieb}, {Grillmair}, {Guhathakurta}, {Han}, {Hou},
  {Lee}, {L{\'e}pine}, {Li}, {Liu}, {Pan}, {Sellwood}, {Wang}, {Wang}, {Yang},
  {Yanny}, {Zhang}, {Zhang}, {Zheng}, \& {Zhu}}]{deng-legue}
{Deng}, L.-C., {Newberg}, H.~J., {Liu}, C., {et~al.} 2012, Research in
  Astronomy and Astrophysics, 12, 735

\bibitem[{{Elsworth} {et~al.}(2017){Elsworth}, {Hekker}, {Basu}, \&
  {Davies}}]{Elsworth2017}
{Elsworth}, Y., {Hekker}, S., {Basu}, S., \& {Davies}, G.~R. 2017, \mnras, 466,
  3344

\bibitem[{{Feuillet} {et~al.}(2018){Feuillet}, {Bovy}, {Holtzman}, {Weinberg},
  {Garc{\'\i}a-Hern{\'a}ndez}, {Hearty}, {Majewski}, {Roman-Lopes}, {Rybizki},
  \& {Zamora}}]{Feuillet2018}
{Feuillet}, D.~K., {Bovy}, J., {Holtzman}, J., {et~al.} 2018, \mnras, 477, 2326

\bibitem[{{Feuillet} {et~al.}(2019){Feuillet}, {Frankel}, {Lind}, {Frinchaboy},
  {Garc{\'\i}a-Hern{\'a}ndez}, {Lane}, {Nitschelm}, \&
  {Roman-Lopes}}]{Feuillet2019}
{Feuillet}, D.~K., {Frankel}, N., {Lind}, K., {et~al.} 2019, \mnras, 489, 1742

\bibitem[{{Gaia Collaboration} {et~al.}(2022){Gaia Collaboration}, {Arenou},
  {Babusiaux}, {Barstow}, {Faigler}, {Jorissen}, {Kervella}, {Mazeh},
  {Mowlavi}, {Panuzzo}, {Sahlmann}, {Shahaf}, {Sozzetti}, {Bauchet},
  {Damerdji}, {Gavras}, {Giacobbe}, {Gosset}, {Halbwachs}, {Holl}, {Lattanzi},
  {Leclerc}, {Morel}, {Pourbaix}, {Re Fiorentin}, {Sadowski}, {S{\'e}gransan},
  {Siopis}, {Teyssier}, {Zwitter}, {Planquart}, {Brown}, {Vallenari}, {Prusti},
  {de Bruijne}, {Biermann}, {Creevey}, {Ducourant}, {Evans}, {Eyer}, {Guerra},
  {Hutton}, {Jordi}, {Klioner}, {Lammers}, {Lindegren}, {Luri}, {Mignard},
  {Panem}, {Randich}, {Sartoretti}, {Soubiran}, {Tanga}, {Walton},
  {Bailer-Jones}, {Bastian}, {Drimmel}, {Jansen}, {Katz}, {van Leeuwen},
  {Bakker}, {Cacciari}, {Casta{\~n}eda}, {De Angeli}, {Fabricius}, {Fouesneau},
  {Fr{\'e}mat}, {Galluccio}, {Guerrier}, {Heiter}, {Masana}, {Messineo},
  {Nicolas}, {Nienartowicz}, {Pailler}, {Riclet}, {Roux}, {Seabroke}, {Sordo},
  {Th{\'e}venin}, {Gracia-Abril}, {Portell}, {Altmann}, {Andrae}, {Audard},
  {Bellas-Velidis}, {Benson}, {Berthier}, {Blomme}, {Burgess}, {Busonero},
  {Busso}, {C{\'a}novas}, {Carry}, {Cellino}, {Cheek}, {Clementini},
  {Davidson}, {de Teodoro}, {Nu{\~n}ez Campos}, {Delchambre}, {Dell'Oro},
  {Esquej}, {Fern{\'a}ndez-Hern{\'a}ndez}, {Fraile}, {Garabato},
  {Garc{\'\i}a-Lario}, {Haigron}, {Hambly}, {Harrison}, {Hern{\'a}ndez},
  {Hestroffer}, {Hodgkin}, {Jan{\ss}en}, {Jevardat de Fombelle}, {Jordan},
  {Krone-Martins}, {Lanzafame}, {L{\"o}ffler}, {Marchal}, {Marrese},
  {Moitinho}, {Muinonen}, {Osborne}, {Pancino}, {Pauwels}, {Recio-Blanco},
  {Reyl{\'e}}, {Riello}, {Rimoldini}, {Roegiers}, {Rybizki}, {Sarro}, {Smith},
  {Utrilla}, {van Leeuwen}, {Abbas}, {{\'A}brah{\'a}m}, {Abreu Aramburu},
  {Aerts}, {Aguado}, {Ajaj}, {Aldea-Montero}, {Altavilla}, {{\'A}lvarez},
  {Alves}, {Anders}, {Anderson}, {Anglada Varela}, {Antoja}, {Baines}, {Baker},
  {Balaguer-N{\'u}{\~n}ez}, {Balbinot}, {Balog}, {Barache}, {Barbato},
  {Barros}, {Bartolom{\'e}}, {Bassilana}, {Becciani}, {Bellazzini},
  {Berihuete}, {Bernet}, {Bertone}, {Bianchi}, {Binnenfeld}, {Blanco-Cuaresma},
  {Blazere}, {Boch}, {Bombrun}, {Bossini}, {Bouquillon}, {Bragaglia},
  {Bramante}, {Breedt}, {Bressan}, {Brouillet}, {Brugaletta}, {Bucciarelli},
  {Burlacu}, {Butkevich}, {Buzzi}, {Caffau}, {Cancelliere}, {Cantat-Gaudin},
  {Carballo}, {Carlucci}, {Carnerero}, {Carrasco}, {Casamiquela}, {Castellani},
  {Castro-Ginard}, {Chaoul}, {Charlot}, {Chemin}, {Chiaramida}, {Chiavassa},
  {Chornay}, {Comoretto}, {Contursi}, {Cooper}, {Cornez}, {Cowell}, {Crifo},
  {Cropper}, {Crosta}, {Crowley}, {Dafonte}, {Dapergolas}, {David}, {de
  Laverny}, {De Luise}, {De March}, {De Ridder}, {de Souza}, {de Torres}, {del
  Peloso}, {del Pozo}, {Delbo}, {Delgado}, {Delisle}, {Demouchy},
  {Dharmawardena}, {Diakite}, {Diener}, {Distefano}, {Dolding}, {Enke},
  {Fabre}, {Fabrizio}, {Fedorets}, {Fernique}, {Figueras}, {Fournier},
  {Fouron}, {Fragkoudi}, {Gai}, {Garcia-Gutierrez}, {Garcia-Reinaldos},
  {Garc{\'\i}a-Torres}, {Garofalo}, {Gavel}, {Gerlach}, {Geyer}, {Gilmore},
  {Girona}, {Giuffrida}, {Gomel}, {Gomez}, {Gonz{\'a}lez-N{\'u}{\~n}ez},
  {Gonz{\'a}lez-Santamar{\'\i}a}, {Gonz{\'a}lez-Vidal}, {Granvik}, {Guillout},
  {Guiraud}, {Guti{\'e}rrez-S{\'a}nchez}, {Guy}, {Hatzidimitriou}, {Hauser},
  {Haywood}, {Helmer}, {Helmi}, {Sarmiento}, {Hidalgo}, {H{\l}adczuk}, {Hobbs},
  {Holland}, {Huckle}, {Jardine}, {Jasniewicz}, {Jean-Antoine Piccolo},
  {Jim{\'e}nez-Arranz}, {Juaristi Campillo}, {Julbe}, {Karbevska}, {Khanna},
  {Kordopatis}, {Korn}, {K{\'o}sp{\'a}l}, {Kostrzewa-Rutkowska},
  {Kruszy{\'n}ska}, {Kun}, {Laizeau}, {Lambert}, {Lanza}, {Lasne}, {Le
  Campion}, {Lebreton}, {Lebzelter}, {Leccia}, {Lecoeur-Taibi}, {Liao},
  {Licata}, {Lindstr{\o}m}, {Lister}, {Livanou}, {Lobel}, {Lorca}, {Loup},
  {Madrero Pardo}, {Magdaleno Romeo}, {Managau}, {Mann}, {Manteiga},
  {Marchant}, {Marconi}, {Marcos}, {Marcos Santos}, {Mar{\'\i}n Pina},
  {Marinoni}, {Marocco}, {Marshall}, {Polo}, {Mart{\'\i}n-Fleitas}, {Marton},
  {Mary}, {Masip}, {Massari}, {Mastrobuono-Battisti}, {McMillan}, {Messina},
  {Michalik}, {Millar}, {Mints}, {Molina}, {Molinaro}, {Moln{\'a}r}, {Monari},
  {Mongui{\'o}}, {Montegriffo}, {Montero}, {Mor}, {Mora}, {Morbidelli},
  {Morris}, {Muraveva}, {Murphy}, {Musella}, {Nagy}, {Noval}, {Oca{\~n}a},
  {Ogden}, {Ordenovic}, {Osinde}, {Pagani}, {Pagano}, {Palaversa}, {Palicio},
  {Pallas-Quintela}, {Panahi}, {Payne-Wardenaar}, {Pe{\~n}alosa Esteller},
  {Penttil{\"a}}, {Pichon}, {Piersimoni}, {Pineau}, {Plachy}, {Plum}, {Poggio},
  {Pr{\v{s}}a}, {Pulone}, {Racero}, {Ragaini}, {Rainer}, {Raiteri}, {Ramos},
  {Ramos-Lerate}, {Regibo}, {Richards}, {Rios Diaz}, {Ripepi}, {Riva}, {Rix},
  {Rixon}, {Robichon}, {Robin}, {Robin}, {Roelens}, {Rogues}, {Rohrbasser},
  {Romero-G{\'o}mez}, {Rowell}, {Royer}, {Ruz Mieres}, {Rybicki}, {S{\'a}ez
  N{\'u}{\~n}ez}, {Sagrist{\`a} Sell{\'e}s}, {Salguero}, {Samaras}, {Sanchez
  Gimenez}, {Sanna}, {Santove{\~n}a}, {Sarasso}, {Schultheis}, {Sciacca},
  {Segol}, {Segovia}, {Semeux}, {Siddiqui}, {Siebert}, {Siltala}, {Silvelo},
  {Slezak}, {Slezak}, {Smart}, {Snaith}, {Solano}, {Solitro}, {Souami},
  {Souchay}, {Spagna}, {Spina}, {Spoto}, {Steele}, {Steidelm{\"u}ller},
  {Stephenson}, {S{\"u}veges}, {Surdej}, {Szabados}, {Szegedi-Elek}, {Taris},
  {Taylor}, {Teixeira}, {Tolomei}, {Tonello}, {Torra}, {Torra}, {Torralba
  Elipe}, {Trabucchi}, {Tsounis}, {Turon}, {Ulla}, {Unger}, {Vaillant}, {van
  Dillen}, {van Reeven}, {Vanel}, {Vecchiato}, {Viala}, {Vicente}, {Voutsinas},
  {Weiler}, {Wevers}, {Wyrzykowski}, {Yoldas}, {Yvard}, {Zhao}, {Zorec}, \&
  {Zucker}}]{gaiadr3}
{Gaia Collaboration}, {Arenou}, F., {Babusiaux}, C., {et~al.} 2022, arXiv
  e-prints, arXiv:2206.05595

\bibitem[{{Gaia Collaboration} {et~al.}(2020){Gaia Collaboration}, {Brown},
  {Vallenari}, {Prusti}, {de Bruijne}, {Babusiaux}, \& {Biermann}}]{Gaiaedr3}
{Gaia Collaboration}, {Brown}, A.~G.~A., {Vallenari}, A., {et~al.} 2020, arXiv
  e-prints, arXiv:2012.01533

\bibitem[{{Gaia Collaboration} {et~al.}(2016){Gaia Collaboration}, {Prusti},
  {de Bruijne}, {Brown}, {Vallenari}, {Babusiaux}, {Bailer-Jones}, {Bastian},
  {Biermann}, {Evans}, {Eyer}, {Jansen}, {Jordi}, {Klioner}, {Lammers},
  {Lindegren}, {Luri}, {Mignard}, {Milligan}, {Panem}, {Poinsignon},
  {Pourbaix}, {Randich}, {Sarri}, {Sartoretti}, {Siddiqui}, {Soubiran},
  {Valette}, {van Leeuwen}, {Walton}, {Aerts}, {Arenou}, {Cropper}, {Drimmel},
  {H{\o}g}, {Katz}, {Lattanzi}, {O'Mullane}, {Grebel}, {Holland}, {Huc},
  {Passot}, {Bramante}, {Cacciari}, {Casta{\~n}eda}, {Chaoul}, {Cheek}, {De
  Angeli}, {Fabricius}, {Guerra}, {Hern{\'a}ndez}, {Jean-Antoine-Piccolo},
  {Masana}, {Messineo}, {Mowlavi}, {Nienartowicz}, {Ord{\'o}{\~n}ez-Blanco},
  {Panuzzo}, {Portell}, {Richards}, {Riello}, {Seabroke}, {Tanga},
  {Th{\'e}venin}, {Torra}, {Els}, {Gracia-Abril}, {Comoretto},
  {Garcia-Reinaldos}, {Lock}, {Mercier}, {Altmann}, {Andrae}, {Astraatmadja},
  {Bellas-Velidis}, {Benson}, {Berthier}, {Blomme}, {Busso}, {Carry},
  {Cellino}, {Clementini}, {Cowell}, {Creevey}, {Cuypers}, {Davidson}, {De
  Ridder}, {de Torres}, {Delchambre}, {Dell'Oro}, {Ducourant}, {Fr{\'e}mat},
  {Garc{\'\i}a-Torres}, {Gosset}, {Halbwachs}, {Hambly}, {Harrison}, {Hauser},
  {Hestroffer}, {Hodgkin}, {Huckle}, {Hutton}, {Jasniewicz}, {Jordan},
  {Kontizas}, {Korn}, {Lanzafame}, {Manteiga}, {Moitinho}, {Muinonen},
  {Osinde}, {Pancino}, {Pauwels}, {Petit}, {Recio-Blanco}, {Robin}, {Sarro},
  {Siopis}, {Smith}, {Smith}, {Sozzetti}, {Thuillot}, {van Reeven}, {Viala},
  {Abbas}, {Abreu Aramburu}, {Accart}, {Aguado}, {Allan}, {Allasia},
  {Altavilla}, {{\'A}lvarez}, {Alves}, {Anderson}, {Andrei}, {Anglada Varela},
  {Antiche}, {Antoja}, {Ant{\'o}n}, {Arcay}, {Atzei}, {Ayache}, {Bach},
  {Baker}, {Balaguer-N{\'u}{\~n}ez}, {Barache}, {Barata}, {Barbier}, {Barblan},
  {Baroni}, {Barrado y Navascu{\'e}s}, {Barros}, {Barstow}, {Becciani},
  {Bellazzini}, {Bellei}, {Bello Garc{\'\i}a}, {Belokurov}, {Bendjoya},
  {Berihuete}, {Bianchi}, {Bienaym{\'e}}, {Billebaud}, {Blagorodnova},
  {Blanco-Cuaresma}, {Boch}, {Bombrun}, {Borrachero}, {Bouquillon}, {Bourda},
  {Bouy}, {Bragaglia}, {Breddels}, {Brouillet}, {Br{\"u}semeister},
  {Bucciarelli}, {Budnik}, {Burgess}, {Burgon}, {Burlacu}, {Busonero}, {Buzzi},
  {Caffau}, {Cambras}, {Campbell}, {Cancelliere}, {Cantat-Gaudin}, {Carlucci},
  {Carrasco}, {Castellani}, {Charlot}, {Charnas}, {Charvet}, {Chassat},
  {Chiavassa}, {Clotet}, {Cocozza}, {Collins}, {Collins}, {Costigan}, {Crifo},
  {Cross}, {Crosta}, {Crowley}, {Dafonte}, {Damerdji}, {Dapergolas}, {David},
  {David}, {De Cat}, {de Felice}, {de Laverny}, {De Luise}, {De March}, {de
  Martino}, {de Souza}, {Debosscher}, {del Pozo}, {Delbo}, {Delgado},
  {Delgado}, {di Marco}, {Di Matteo}, {Diakite}, {Distefano}, {Dolding}, {Dos
  Anjos}, {Drazinos}, {Dur{\'a}n}, {Dzigan}, {Ecale}, {Edvardsson}, {Enke},
  {Erdmann}, {Escolar}, {Espina}, {Evans}, {Eynard Bontemps}, {Fabre},
  {Fabrizio}, {Faigler}, {Falc{\~a}o}, {Farr{\`a}s Casas}, {Faye}, {Federici},
  {Fedorets}, {Fern{\'a}ndez-Hern{\'a}ndez}, {Fernique}, {Fienga}, {Figueras},
  {Filippi}, {Findeisen}, {Fonti}, {Fouesneau}, {Fraile}, {Fraser}, {Fuchs},
  {Furnell}, {Gai}, {Galleti}, {Galluccio}, {Garabato}, {Garc{\'\i}a-Sedano},
  {Gar{\'e}}, {Garofalo}, {Garralda}, {Gavras}, {Gerssen}, {Geyer}, {Gilmore},
  {Girona}, {Giuffrida}, {Gomes}, {Gonz{\'a}lez-Marcos},
  {Gonz{\'a}lez-N{\'u}{\~n}ez}, {Gonz{\'a}lez-Vidal}, {Granvik}, {Guerrier},
  {Guillout}, {Guiraud}, {G{\'u}rpide}, {Guti{\'e}rrez-S{\'a}nchez}, {Guy},
  {Haigron}, {Hatzidimitriou}, {Haywood}, {Heiter}, {Helmi}, {Hobbs},
  {Hofmann}, {Holl}, {Holland}, {Hunt}, {Hypki}, {Icardi}, {Irwin}, {Jevardat
  de Fombelle}, {Jofr{\'e}}, {Jonker}, {Jorissen}, {Julbe}, {Karampelas},
  {Kochoska}, {Kohley}, {Kolenberg}, {Kontizas}, {Koposov}, {Kordopatis},
  {Koubsky}, {Kowalczyk}, {Krone-Martins}, {Kudryashova}, {Kull}, {Bachchan},
  {Lacoste-Seris}, {Lanza}, {Lavigne}, {Le Poncin-Lafitte}, {Lebreton},
  {Lebzelter}, {Leccia}, {Leclerc}, {Lecoeur-Taibi}, {Lemaitre}, {Lenhardt},
  {Leroux}, {Liao}, {Licata}, {Lindstr{\o}m}, {Lister}, {Livanou}, {Lobel},
  {L{\"o}ffler}, {L{\'o}pez}, {Lopez-Lozano}, {Lorenz}, {Loureiro},
  {MacDonald}, {Magalh{\~a}es Fernandes}, {Managau}, {Mann}, {Mantelet},
  {Marchal}, {Marchant}, {Marconi}, {Marie}, {Marinoni}, {Marrese},
  {Marschalk{\'o}}, {Marshall}, {Mart{\'\i}n-Fleitas}, {Martino}, {Mary},
  {Matijevi{\v{c}}}, {Mazeh}, {McMillan}, {Messina}, {Mestre}, {Michalik},
  {Millar}, {Miranda}, {Molina}, {Molinaro}, {Molinaro}, {Moln{\'a}r},
  {Moniez}, {Montegriffo}, {Monteiro}, {Mor}, {Mora}, {Morbidelli}, {Morel},
  {Morgenthaler}, {Morley}, {Morris}, {Mulone}, {Muraveva}, {Musella},
  {Narbonne}, {Nelemans}, {Nicastro}, {Noval}, {Ord{\'e}novic},
  {Ordieres-Mer{\'e}}, {Osborne}, {Pagani}, {Pagano}, {Pailler}, {Palacin},
  {Palaversa}, {Parsons}, {Paulsen}, {Pecoraro}, {Pedrosa}, {Pentik{\"a}inen},
  {Pereira}, {Pichon}, {Piersimoni}, {Pineau}, {Plachy}, {Plum}, {Poujoulet},
  {Pr{\v{s}}a}, {Pulone}, {Ragaini}, {Rago}, {Rambaux}, {Ramos-Lerate},
  {Ranalli}, {Rauw}, {Read}, {Regibo}, {Renk}, {Reyl{\'e}}, {Ribeiro},
  {Rimoldini}, {Ripepi}, {Riva}, {Rixon}, {Roelens}, {Romero-G{\'o}mez},
  {Rowell}, {Royer}, {Rudolph}, {Ruiz-Dern}, {Sadowski}, {Sagrist{\`a}
  Sell{\'e}s}, {Sahlmann}, {Salgado}, {Salguero}, {Sarasso}, {Savietto},
  {Schnorhk}, {Schultheis}, {Sciacca}, {Segol}, {Segovia}, {Segransan},
  {Serpell}, {Shih}, {Smareglia}, {Smart}, {Smith}, {Solano}, {Solitro},
  {Sordo}, {Soria Nieto}, {Souchay}, {Spagna}, {Spoto}, {Stampa}, {Steele},
  {Steidelm{\"u}ller}, {Stephenson}, {Stoev}, {Suess}, {S{\"u}veges}, {Surdej},
  {Szabados}, {Szegedi-Elek}, {Tapiador}, {Taris}, {Tauran}, {Taylor},
  {Teixeira}, {Terrett}, {Tingley}, {Trager}, {Turon}, {Ulla}, {Utrilla},
  {Valentini}, {van Elteren}, {Van Hemelryck}, {van Leeuwen}, {Varadi},
  {Vecchiato}, {Veljanoski}, {Via}, {Vicente}, {Vogt}, {Voss}, {Votruba},
  {Voutsinas}, {Walmsley}, {Weiler}, {Weingrill}, {Werner}, {Wevers},
  {Whitehead}, {Wyrzykowski}, {Yoldas}, {{\v{Z}}erjal}, {Zucker}, {Zurbach},
  {Zwitter}, {Alecu}, {Allen}, {Allende Prieto}, {Amorim},
  {Anglada-Escud{\'e}}, {Arsenijevic}, {Azaz}, {Balm}, {Beck}, {Bernstein},
  {Bigot}, {Bijaoui}, {Blasco}, {Bonfigli}, {Bono}, {Boudreault}, {Bressan},
  {Brown}, {Brunet}, {Bunclark}, {Buonanno}, {Butkevich}, {Carret}, {Carrion},
  {Chemin}, {Ch{\'e}reau}, {Corcione}, {Darmigny}, {de Boer}, {de Teodoro}, {de
  Zeeuw}, {Delle Luche}, {Domingues}, {Dubath}, {Fodor}, {Fr{\'e}zouls},
  {Fries}, {Fustes}, {Fyfe}, {Gallardo}, {Gallegos}, {Gardiol}, {Gebran},
  {Gomboc}, {G{\'o}mez}, {Grux}, {Gueguen}, {Heyrovsky}, {Hoar}, {Iannicola},
  {Isasi Parache}, {Janotto}, {Joliet}, {Jonckheere}, {Keil}, {Kim},
  {Klagyivik}, {Klar}, {Knude}, {Kochukhov}, {Kolka}, {Kos}, {Kutka}, {Lainey},
  {LeBouquin}, {Liu}, {Loreggia}, {Makarov}, {Marseille}, {Martayan},
  {Martinez-Rubi}, {Massart}, {Meynadier}, {Mignot}, {Munari}, {Nguyen},
  {Nordlander}, {Ocvirk}, {O'Flaherty}, {Olias Sanz}, {Ortiz}, {Osorio},
  {Oszkiewicz}, {Ouzounis}, {Palmer}, {Park}, {Pasquato}, {Peltzer}, {Peralta},
  {P{\'e}turaud}, {Pieniluoma}, {Pigozzi}, {Poels}, {Prat}, {Prod'homme},
  {Raison}, {Rebordao}, {Risquez}, {Rocca-Volmerange}, {Rosen}, {Ruiz-Fuertes},
  {Russo}, {Sembay}, {Serraller Vizcaino}, {Short}, {Siebert}, {Silva},
  {Sinachopoulos}, {Slezak}, {Soffel}, {Sosnowska}, {Strai{\v{z}}ys}, {ter
  Linden}, {Terrell}, {Theil}, {Tiede}, {Troisi}, {Tsalmantza}, {Tur},
  {Vaccari}, {Vachier}, {Valles}, {Van Hamme}, {Veltz}, {Virtanen}, {Wallut},
  {Wichmann}, {Wilkinson}, {Ziaeepour}, \& {Zschocke}}]{gaiamission}
{Gaia Collaboration}, {Prusti}, T., {de Bruijne}, J.~H.~J., {et~al.} 2016,
  \aap, 595, A1

\bibitem[{{Hasselquist} {et~al.}(2019){Hasselquist}, {Holtzman}, {Shetrone},
  {Tayar}, {Weinberg}, {Feuillet}, {Cunha}, {Pinsonneault}, {Johnson}, {Bird},
  {Beers}, {Schiavon}, {Minchev}, {Fern{\'a}ndez-Trincado},
  {Garc{\'\i}a-Hern{\'a}ndez}, {Nitschelm}, \& {Zamora}}]{Hasselquist2019}
{Hasselquist}, S., {Holtzman}, J.~A., {Shetrone}, M., {et~al.} 2019, \apj, 871,
  181

\bibitem[{{Hawkins} {et~al.}(2018){Hawkins}, {Ting}, \&
  {Walter-Rix}}]{Hawkins2018}
{Hawkins}, K., {Ting}, Y.-S., \& {Walter-Rix}, H. 2018, \apj, 853, 20

\bibitem[{{Hayden} {et~al.}(2015){Hayden}, {Bovy}, {Holtzman}, {Nidever},
  {Bird}, {Weinberg}, {Andrews}, {Majewski}, {Allende Prieto}, {Anders},
  {Beers}, {Bizyaev}, {Chiappini}, {Cunha}, {Frinchaboy},
  {Garc{\'\i}a-Her{\'n}andez}, {Garc{\'\i}a P{\'e}rez}, {Girardi}, {Harding},
  {Hearty}, {Johnson}, {M{\'e}sz{\'a}ros}, {Minchev}, {O'Connell}, {Pan},
  {Robin}, {Schiavon}, {Schneider}, {Schultheis}, {Shetrone}, {Skrutskie},
  {Steinmetz}, {Smith}, {Wilson}, {Zamora}, \& {Zasowski}}]{Hayden2015}
{Hayden}, M.~R., {Bovy}, J., {Holtzman}, J.~A., {et~al.} 2015, \apj, 808, 132

\bibitem[{{Haywood}(2008)}]{Haywood2008}
{Haywood}, M. 2008, \mnras, 388, 1175

\bibitem[{{Haywood} {et~al.}(2013){Haywood}, {Di Matteo}, {Lehnert}, {Katz}, \&
  {G{\'o}mez}}]{Haywood2013}
{Haywood}, M., {Di Matteo}, P., {Lehnert}, M.~D., {Katz}, D., \& {G{\'o}mez},
  A. 2013, \aap, 560, A109

\bibitem[{{Ho} {et~al.}(2017){Ho}, {Ness}, {Hogg}, {Rix}, {Liu}, {Yang},
  {Zhang}, {Hou}, \& {Wang}}]{Ho2017}
{Ho}, A. Y.~Q., {Ness}, M.~K., {Hogg}, D.~W., {et~al.} 2017, \apj, 836, 5

\bibitem[{{Huang} {et~al.}(2015){Huang}, {Liu}, {Zhang}, {Yuan}, {Xiang},
  {Chen}, {Ren}, {Sun}, {Wang}, {Zhang}, {Hou}, {Wang}, \& {Yang}}]{Huang2015}
{Huang}, Y., {Liu}, X.-W., {Zhang}, H.-W., {et~al.} 2015, Research in Astronomy
  and Astrophysics, 15, 1240

\bibitem[{{Huang} {et~al.}(2020){Huang}, {Sch{\"o}nrich}, {Zhang}, {Wu},
  {Chen}, {Wang}, {Xiang}, {Wang}, {Yuan}, {Li}, {Sun}, {Li}, \&
  {Liu}}]{Huang2020}
{Huang}, Y., {Sch{\"o}nrich}, R., {Zhang}, H., {et~al.} 2020, \apjs, 249, 29

\bibitem[{{Huber} {et~al.}(2011){Huber}, {Bedding}, {Stello}, {Hekker},
  {Mathur}, {Mosser}, {Verner}, {Bonanno}, {Buzasi}, {Campante}, {Elsworth},
  {Hale}, {Kallinger}, {Silva Aguirre}, {Chaplin}, {De Ridder}, {Garc{\'\i}a},
  {Appourchaux}, {Frandsen}, {Houdek}, {Molenda-{\.Z}akowicz}, {Monteiro},
  {Christensen-Dalsgaard}, {Gilliland}, {Kawaler}, {Kjeldsen}, {Broomhall},
  {Corsaro}, {Salabert}, {Sanderfer}, {Seader}, \& {Smith}}]{Huber2011}
{Huber}, D., {Bedding}, T.~R., {Stello}, D., {et~al.} 2011, \apj, 743, 143

\bibitem[{{Huber} {et~al.}(2014){Huber}, {Silva Aguirre}, {Matthews},
  {Pinsonneault}, {Gaidos}, {Garc{\'\i}a}, {Hekker}, {Mathur}, {Mosser},
  {Torres}, {Bastien}, {Basu}, {Bedding}, {Chaplin}, {Demory}, {Fleming},
  {Guo}, {Mann}, {Rowe}, {Serenelli}, {Smith}, \& {Stello}}]{Huber2014}
{Huber}, D., {Silva Aguirre}, V., {Matthews}, J.~M., {et~al.} 2014, \apjs, 211,
  2

\bibitem[{{Kharchenko} {et~al.}(2013){Kharchenko}, {Piskunov}, {Schilbach},
  {R{\"o}ser}, \& {Scholz}}]{Kharchenko2013}
{Kharchenko}, N.~V., {Piskunov}, A.~E., {Schilbach}, E., {R{\"o}ser}, S., \&
  {Scholz}, R.~D. 2013, \aap, 558, A53

\bibitem[{{Lebreton} \& {Goupil}(2014)}]{Lebreton2014}
{Lebreton}, Y. \& {Goupil}, M.~J. 2014, \aap, 569, A21

\bibitem[{{Li} {et~al.}(2020){Li}, {Bedding}, {Christensen-Dalsgaard},
  {Stello}, {Li}, \& {Keen}}]{Li2020}
{Li}, T., {Bedding}, T.~R., {Christensen-Dalsgaard}, J., {et~al.} 2020, \mnras,
  495, 3431

\bibitem[{{Li} {et~al.}(2022{\natexlab{a}}){Li}, {Li}, {Bi}, {Bedding},
  {Davies}, \& {Du}}]{Lit2022}
{Li}, T., {Li}, Y., {Bi}, S., {et~al.} 2022{\natexlab{a}}, \apj, 927, 167

\bibitem[{{Li} {et~al.}(2022{\natexlab{b}}){Li}, {Bedding}, {Murphy}, {Stello},
  {Chen}, {Huber}, {Joyce}, {Marks}, {Zhang}, {Bi}, {Colman}, {Hayden}, {Hey},
  {Li}, {Montet}, {Sharma}, \& {Wu}}]{Li2022}
{Li}, Y., {Bedding}, T.~R., {Murphy}, S.~J., {et~al.} 2022{\natexlab{b}},
  Nature Astronomy, 6, 673

\bibitem[{{Liu} {et~al.}(2014){Liu}, {Yuan}, {Huo}, {Deng}, {Hou}, {Zhao},
  {Zhao}, {Shi}, {Luo}, {Xiang}, {Zhang}, {Huang}, \& {Zhang}}]{liu-lss-gac}
{Liu}, X.-W., {Yuan}, H.-B., {Huo}, Z.-Y., {et~al.} 2014, in IAU Symposium,
  Vol. 298, IAU Symposium, ed. S.~{Feltzing}, G.~{Zhao}, N.~A. {Walton}, \&
  P.~{Whitelock}, 310--321

\bibitem[{{Luo} {et~al.}(2015){Luo}, {Zhao}, {Zhao}, {Deng}, {Liu}, {Jing},
  {Wang}, {Zhang}, {Shi}, {Cui}, {Chu}, {Li}, {Bai}, {Wu}, {Cai}, {Cao}, {Cao},
  {Carlin}, {Chen}, {Chen}, {Chen}, {Chen}, {Chen}, {Chen}, {Chen},
  {Christlieb}, {Chu}, {Cui}, {Dong}, {Du}, {Fan}, {Feng}, {Fu}, {Gao}, {Gong},
  {Gu}, {Guo}, {Han}, {He}, {Hou}, {Hou}, {Hou}, {Hu}, {Hu}, {Hu}, {Huo},
  {Jia}, {Jiang}, {Jiang}, {Jiang}, {Jin}, {Kong}, {Kong}, {Lei}, {Li}, {Li},
  {Li}, {Li}, {Li}, {Li}, {Li}, {Li}, {Li}, {Li}, {Li}, {Li}, {Liang}, {Lin},
  {Liu}, {Liu}, {Liu}, {Liu}, {Lu}, {Luo}, {Mao}, {Newberg}, {Ni}, {Qi}, {Qi},
  {Shen}, {Shi}, {Song}, {Song}, {Su}, {Su}, {Tang}, {Tao}, {Tian}, {Wang},
  {Wang}, {Wang}, {Wang}, {Wang}, {Wang}, {Wang}, {Wang}, {Wang}, {Wang},
  {Wang}, {Wang}, {Wang}, {Wang}, {Wang}, {Wang}, {Wang}, {Wang}, {Wang},
  {Wang}, {Wei}, {Wei}, {Wu}, {Wu}, {Wu}, {Wu}, {Xing}, {Xu}, {Xu}, {Xu},
  {Yan}, {Yang}, {Yang}, {Yang}, {Yang}, {Yao}, {Yu}, {Yuan}, {Yuan}, {Yuan},
  {Yuan}, {Zhai}, {Zhang}, {Zhang}, {Zhang}, {Zhang}, {Zhang}, {Zhang},
  {Zhang}, {Zhang}, {Zhao}, {Zhou}, {Zhou}, {Zhu}, {Zhu}, {Zou}, \&
  {Zuo}}]{luo2015}
{Luo}, A.~L., {Zhao}, Y.-H., {Zhao}, G., {et~al.} 2015, Research in Astronomy
  and Astrophysics, 15, 1095

\bibitem[{{Majewski} {et~al.}(2017){Majewski}, {Schiavon}, {Frinchaboy},
  {Allende Prieto}, {Barkhouser}, {Bizyaev}, {Blank}, {Brunner}, {Burton},
  {Carrera}, {Chojnowski}, {Cunha}, {Epstein}, {Fitzgerald}, {Garc{\'\i}a
  P{\'e}rez}, {Hearty}, {Henderson}, {Holtzman}, {Johnson}, {Lam}, {Lawler},
  {Maseman}, {M{\'e}sz{\'a}ros}, {Nelson}, {Nguyen}, {Nidever}, {Pinsonneault},
  {Shetrone}, {Smee}, {Smith}, {Stolberg}, {Skrutskie}, {Walker}, {Wilson},
  {Zasowski}, {Anders}, {Basu}, {Beland}, {Blanton}, {Bovy}, {Brownstein},
  {Carlberg}, {Chaplin}, {Chiappini}, {Eisenstein}, {Elsworth}, {Feuillet},
  {Fleming}, {Galbraith-Frew}, {Garc{\'\i}a}, {Garc{\'\i}a-Hern{\'a}ndez},
  {Gillespie}, {Girardi}, {Gunn}, {Hasselquist}, {Hayden}, {Hekker}, {Ivans},
  {Kinemuchi}, {Klaene}, {Mahadevan}, {Mathur}, {Mosser}, {Muna}, {Munn},
  {Nichol}, {O'Connell}, {Parejko}, {Robin}, {Rocha-Pinto}, {Schultheis},
  {Serenelli}, {Shane}, {Silva Aguirre}, {Sobeck}, {Thompson}, {Troup},
  {Weinberg}, \& {Zamora}}]{Majewski2017}
{Majewski}, S.~R., {Schiavon}, R.~P., {Frinchaboy}, P.~M., {et~al.} 2017, \aj,
  154, 94

\bibitem[{{Martell} {et~al.}(2008){Martell}, {Smith}, \&
  {Briley}}]{Martell2008}
{Martell}, S.~L., {Smith}, G.~H., \& {Briley}, M.~M. 2008, \aj, 136, 2522

\bibitem[{{Martig} {et~al.}(2016{\natexlab{a}}){Martig}, {Fouesneau}, {Rix},
  {Ness}, {M{\'e}sz{\'a}ros}, {Garc{\'\i}a-Hern{\'a}ndez}, {Pinsonneault},
  {Serenelli}, {Silva Aguirre}, \& {Zamora}}]{Martig2016a}
{Martig}, M., {Fouesneau}, M., {Rix}, H.-W., {et~al.} 2016{\natexlab{a}},
  \mnras, 456, 3655

\bibitem[{{Martig} {et~al.}(2016{\natexlab{b}}){Martig}, {Minchev}, {Ness},
  {Fouesneau}, \& {Rix}}]{Martig2016}
{Martig}, M., {Minchev}, I., {Ness}, M., {Fouesneau}, M., \& {Rix}, H.-W.
  2016{\natexlab{b}}, \apj, 831, 139

\bibitem[{{Masseron} \& {Gilmore}(2015)}]{Masseron2015}
{Masseron}, T. \& {Gilmore}, G. 2015, \mnras, 453, 1855

\bibitem[{{Masseron} \& {Hawkins}(2017)}]{Masseron2017b}
{Masseron}, T. \& {Hawkins}, K. 2017, \aap, 597, L3

\bibitem[{{Masseron} {et~al.}(2017){Masseron}, {Lagarde}, {Miglio}, {Elsworth},
  \& {Gilmore}}]{Masseron2017a}
{Masseron}, T., {Lagarde}, N., {Miglio}, A., {Elsworth}, Y., \& {Gilmore}, G.
  2017, \mnras, 464, 3021

\bibitem[{{Miglio} {et~al.}(2017){Miglio}, {Chiappini}, {Mosser}, {Davies},
  {Freeman}, {Girardi}, {Jofr{\'e}}, {Kawata}, {Rendle}, {Valentini},
  {Casagrande}, {Chaplin}, {Gilmore}, {Hawkins}, {Holl}, {Appourchaux},
  {Belkacem}, {Bossini}, {Brogaard}, {Goupil}, {Montalb{\'a}n}, {Noels},
  {Anders}, {Rodrigues}, {Piotto}, {Pollacco}, {Rauer}, {Allende Prieto},
  {Avelino}, {Babusiaux}, {Barban}, {Barbuy}, {Basu}, {Baudin}, {Benomar},
  {Bienaym{\'e}}, {Binney}, {Bland-Hawthorn}, {Bressan}, {Cacciari},
  {Campante}, {Cassisi}, {Christensen-Dalsgaard}, {Combes}, {Creevey}, {Cunha},
  {Jong}, {Laverny}, {Degl'Innocenti}, {Deheuvels}, {Depagne}, {Ridder},
  {Matteo}, {Mauro}, {Dupret}, {Eggenberger}, {Elsworth}, {Famaey}, {Feltzing},
  {Garc{\'\i}a}, {Gerhard}, {Gibson}, {Gizon}, {Haywood}, {Handberg}, {Heiter},
  {Hekker}, {Huber}, {Ibata}, {Katz}, {Kawaler}, {Kjeldsen}, {Kurtz},
  {Lagarde}, {Lebreton}, {Lund}, {Majewski}, {Marigo}, {Martig}, {Mathur},
  {Minchev}, {Morel}, {Ortolani}, {Pinsonneault}, {Plez}, {Moroni}, {Pricopi},
  {Recio-Blanco}, {Reyl{\'e}}, {Robin}, {Roxburgh}, {Salaris}, {Santiago},
  {Schiavon}, {Serenelli}, {Sharma}, {Aguirre}, {Soubiran}, {Steinmetz},
  {Stello}, {Strassmeier}, {Ventura}, {Ventura}, {Walton}, \&
  {Worley}}]{Miglio2017}
{Miglio}, A., {Chiappini}, C., {Mosser}, B., {et~al.} 2017, Astronomische
  Nachrichten, 338, 644

\bibitem[{{Molenda-{\.Z}akowicz} {et~al.}(2014){Molenda-{\.Z}akowicz},
  {Brogaard}, {Niemczura}, {Bergemann}, {Frasca}, {Arentoft}, \&
  {Grundahl}}]{Molenda2014}
{Molenda-{\.Z}akowicz}, J., {Brogaard}, K., {Niemczura}, E., {et~al.} 2014,
  \mnras, 445, 2446

\bibitem[{{Montalb{\'a}n} {et~al.}(2021){Montalb{\'a}n}, {Mackereth}, {Miglio},
  {Vincenzo}, {Chiappini}, {Buldgen}, {Mosser}, {Noels}, {Scuflaire}, {Vrard},
  {Willett}, {Davies}, {Hall}, {Nielsen}, {Khan}, {Rendle}, {van Rossem},
  {Ferguson}, \& {Chaplin}}]{Montalban2021}
{Montalb{\'a}n}, J., {Mackereth}, J.~T., {Miglio}, A., {et~al.} 2021, Nature
  Astronomy, 5, 640

\bibitem[{{Montalb{\'a}n} {et~al.}(2010){Montalb{\'a}n}, {Miglio}, {Noels},
  {Scuflaire}, \& {Ventura}}]{Montalb2010}
{Montalb{\'a}n}, J., {Miglio}, A., {Noels}, A., {Scuflaire}, R., \& {Ventura},
  P. 2010, \apjl, 721, L182

\bibitem[{{Mosser} {et~al.}(2011){Mosser}, {Barban}, {Montalb{\'a}n}, {Beck},
  {Miglio}, {Belkacem}, {Goupil}, {Hekker}, {De Ridder}, {Dupret}, {Elsworth},
  {Noels}, {Baudin}, {Michel}, {Samadi}, {Auvergne}, {Baglin}, \&
  {Catala}}]{Mosser2011}
{Mosser}, B., {Barban}, C., {Montalb{\'a}n}, J., {et~al.} 2011, \aap, 532, A86

\bibitem[{{Mosser} {et~al.}(2012){Mosser}, {Goupil}, {Belkacem}, {Michel},
  {Stello}, {Marques}, {Elsworth}, {Barban}, {Beck}, {Bedding}, {De Ridder},
  {Garc{\'\i}a}, {Hekker}, {Kallinger}, {Samadi}, {Stumpe}, {Barclay}, \&
  {Burke}}]{Mosser2012}
{Mosser}, B., {Goupil}, M.~J., {Belkacem}, K., {et~al.} 2012, \aap, 540, A143

\bibitem[{{Ness} {et~al.}(2016){Ness}, {Hogg}, {Rix}, {Martig}, {Pinsonneault},
  \& {Ho}}]{Ness2016}
{Ness}, M., {Hogg}, D.~W., {Rix}, H.~W., {et~al.} 2016, \apj, 823, 114

\bibitem[{{Pinsonneault} {et~al.}(2014){Pinsonneault}, {Elsworth}, {Epstein},
  {Hekker}, {M{\'e}sz{\'a}ros}, {Chaplin}, {Johnson}, {Garc{\'\i}a},
  {Holtzman}, {Mathur}, {Garc{\'\i}a P{\'e}rez}, {Silva Aguirre}, {Girardi},
  {Basu}, {Shetrone}, {Stello}, {Allende Prieto}, {An}, {Beck}, {Beers},
  {Bizyaev}, {Bloemen}, {Bovy}, {Cunha}, {De Ridder}, {Frinchaboy},
  {Garc{\'\i}a-Hern{\'a}ndez}, {Gilliland}, {Harding}, {Hearty}, {Huber},
  {Ivans}, {Kallinger}, {Majewski}, {Metcalfe}, {Miglio}, {Mosser}, {Muna},
  {Nidever}, {Schneider}, {Serenelli}, {Smith}, {Tayar}, {Zamora}, \&
  {Zasowski}}]{Pinsonneault2014}
{Pinsonneault}, M.~H., {Elsworth}, Y., {Epstein}, C., {et~al.} 2014, \apjs,
  215, 19

\bibitem[{{Rendle} {et~al.}(2019){Rendle}, {Buldgen}, {Miglio}, {Reese},
  {Noels}, {Davies}, {Campante}, {Chaplin}, {Lund}, {Kuszlewicz}, {Scott},
  {Scuflaire}, {Ball}, {Smetana}, \& {Nsamba}}]{Rendle2019}
{Rendle}, B.~M., {Buldgen}, G., {Miglio}, A., {et~al.} 2019, \mnras, 484, 771

\bibitem[{{Salaris} {et~al.}(2004){Salaris}, {Weiss}, \&
  {Percival}}]{Salaris2004}
{Salaris}, M., {Weiss}, A., \& {Percival}, S.~M. 2004, \aap, 414, 163

\bibitem[{{Sanders} \& {Das}(2018)}]{Sanders2018}
{Sanders}, J.~L. \& {Das}, P. 2018, \mnras, 481, 4093

\bibitem[{{Sandquist} {et~al.}(2016){Sandquist}, {Jessen-Hansen}, {Shetrone},
  {Brogaard}, {Meibom}, {Leitner}, {Stello}, {Bruntt}, {Antoci}, {Orosz},
  {Grundahl}, \& {Frandsen}}]{Sandquist2016}
{Sandquist}, E.~L., {Jessen-Hansen}, J., {Shetrone}, M.~D., {et~al.} 2016,
  \apj, 831, 11

\bibitem[{{Sharma} {et~al.}(2016){Sharma}, {Stello}, {Bland-Hawthorn}, {Huber},
  \& {Bedding}}]{Sharma2016}
{Sharma}, S., {Stello}, D., {Bland-Hawthorn}, J., {Huber}, D., \& {Bedding},
  T.~R. 2016, \apj, 822, 15

\bibitem[{{Soderblom}(2010)}]{Soderblom2010}
{Soderblom}, D.~R. 2010, \araa, 48, 581

\bibitem[{{Steinmetz} {et~al.}(2006){Steinmetz}, {Zwitter}, {Siebert},
  {Watson}, {Freeman}, {Munari}, {Campbell}, {Williams}, {Seabroke}, {Wyse},
  {Parker}, {Bienaym{\'e}}, {Roeser}, {Gibson}, {Gilmore}, {Grebel}, {Helmi},
  {Navarro}, {Burton}, {Cass}, {Dawe}, {Fiegert}, {Hartley}, {Russell},
  {Saunders}, {Enke}, {Bailin}, {Binney}, {Bland-Hawthorn}, {Boeche}, {Dehnen},
  {Eisenstein}, {Evans}, {Fiorucci}, {Fulbright}, {Gerhard}, {Jauregi}, {Kelz},
  {Mijovi{\'c}}, {Minchev}, {Parmentier}, {Pe{\~n}arrubia}, {Quillen}, {Read},
  {Ruchti}, {Scholz}, {Siviero}, {Smith}, {Sordo}, {Veltz}, {Vidrih}, {von
  Berlepsch}, {Boyle}, \& {Schilbach}}]{Steinmetz2006}
{Steinmetz}, M., {Zwitter}, T., {Siebert}, A., {et~al.} 2006, \aj, 132, 1645

\bibitem[{{Stello} {et~al.}(2013){Stello}, {Huber}, {Bedding}, {Benomar},
  {Bildsten}, {Elsworth}, {Gilliland}, {Mosser}, {Paxton}, \&
  {White}}]{Stello2013}
{Stello}, D., {Huber}, D., {Bedding}, T.~R., {et~al.} 2013, \apjl, 765, L41

\bibitem[{{Stello} {et~al.}(2016){Stello}, {Vanderburg}, {Casagrande},
  {Gilliland}, {Silva Aguirre}, {Sandquist}, {Leiner}, {Mathieu}, \&
  {Soderblom}}]{Stello2016}
{Stello}, D., {Vanderburg}, A., {Casagrande}, L., {et~al.} 2016, \apj, 832, 133

\bibitem[{{Tayar} {et~al.}(2017){Tayar}, {Somers}, {Pinsonneault}, {Stello},
  {Mints}, {Johnson}, {Zamora}, {Garc{\'\i}a-Hern{\'a}ndez}, {Maraston},
  {Serenelli}, {Allende Prieto}, {Bastien}, {Basu}, {Bird}, {Cohen}, {Cunha},
  {Elsworth}, {Garc{\'\i}a}, {Girardi}, {Hekker}, {Holtzman}, {Huber},
  {Mathur}, {M{\'e}sz{\'a}ros}, {Mosser}, {Shetrone}, {Silva Aguirre},
  {Stassun}, {Stringfellow}, {Zasowski}, \& {Roman-Lopes}}]{Tayar2017}
{Tayar}, J., {Somers}, G., {Pinsonneault}, M.~H., {et~al.} 2017, \apj, 840, 17

\bibitem[{{Ting} {et~al.}(2018){Ting}, {Hawkins}, \& {Rix}}]{Ting2018}
{Ting}, Y.-S., {Hawkins}, K., \& {Rix}, H.-W. 2018, \apjl, 858, L7

\bibitem[{{Vrard} {et~al.}(2016){Vrard}, {Mosser}, \& {Samadi}}]{Vrard2016}
{Vrard}, M., {Mosser}, B., \& {Samadi}, R. 2016, \aap, 588, A87

\bibitem[{{Wang} {et~al.}(2022){Wang}, {Huang}, {Yuan}, {Zhang}, {Xiang}, \&
  {Liu}}]{Wang2022}
{Wang}, C., {Huang}, Y., {Yuan}, H., {et~al.} 2022, \apjs, 259, 51

\bibitem[{{Wang} {et~al.}(2019){Wang}, {Liu}, {Xiang}, {Huang}, {Chen}, {Yuan},
  {Ren}, {Zhang}, \& {Tian}}]{Wang2019}
{Wang}, C., {Liu}, X.~W., {Xiang}, M.~S., {et~al.} 2019, \mnras, 482, 2189

\bibitem[{{Wu} {et~al.}(2018){Wu}, {Xiang}, {Bi}, {Liu}, {Yu}, {Hon}, {Sharma},
  {Li}, {Huang}, {Liu}, {Zhang}, {Li}, {Ge}, {Tian}, {Zhang}, \&
  {Zhang}}]{Wu2018}
{Wu}, Y., {Xiang}, M., {Bi}, S., {et~al.} 2018, \mnras, 475, 3633

\bibitem[{{Wu} {et~al.}(2019){Wu}, {Xiang}, {Zhao}, {Bi}, {Liu}, {Shi},
  {Huang}, {Yuan}, {Wang}, {Chen}, {Huo}, {Ren}, {Tian}, {Liu}, {Zhang}, {Li},
  \& {Zhang}}]{Wuyaqian2019}
{Wu}, Y., {Xiang}, M., {Zhao}, G., {et~al.} 2019, \mnras, 484, 5315

\bibitem[{{Wu} {et~al.}(2023){Wu}, {Xiang}, {Zhao}, {Chen}, {Bi}, \&
  {Li}}]{Wu2023}
{Wu}, Y., {Xiang}, M., {Zhao}, G., {et~al.} 2023, \mnras, 520, 1913

\bibitem[{{Xiang} {et~al.}(2017){Xiang}, {Liu}, {Shi}, {Yuan}, {Huang}, {Chen},
  {Wang}, {Tian}, {Wu}, {Yang}, {Zhang}, {Huo}, \& {Ren}}]{Xiang2017}
{Xiang}, M., {Liu}, X., {Shi}, J., {et~al.} 2017, \apjs, 232, 2

\bibitem[{{Xiang} {et~al.}(2015){Xiang}, {Liu}, {Yuan}, {Huang}, {Wang}, {Ren},
  {Chen}, {Sun}, {Zhang}, {Huo}, \& {Rebassa-Mansergas}}]{Xiangmdf}
{Xiang}, M.-S., {Liu}, X.-W., {Yuan}, H.-B., {et~al.} 2015, Research in
  Astronomy and Astrophysics, 15, 1209

\bibitem[{{Yang} {et~al.}(2012){Yang}, {Meng}, {Bi}, {Tian}, {Liu}, {Li}, \&
  {Li}}]{Yang2012}
{Yang}, W., {Meng}, X., {Bi}, S., {et~al.} 2012, \mnras, 422, 1552

\bibitem[{{Yanny} {et~al.}(2009){Yanny}, {Rockosi}, {Newberg}, {Knapp},
  {Adelman-McCarthy}, {Alcorn}, {Allam}, {Allende Prieto}, {An}, {Anderson},
  {Anderson}, {Bailer-Jones}, {Bastian}, {Beers}, {Bell}, {Belokurov},
  {Bizyaev}, {Blythe}, {Bochanski}, {Boroski}, {Brinchmann}, {Brinkmann},
  {Brewington}, {Carey}, {Cudworth}, {Evans}, {Evans}, {Gates}, {G{\"a}nsicke},
  {Gillespie}, {Gilmore}, {Nebot Gomez-Moran}, {Grebel}, {Greenwell}, {Gunn},
  {Jordan}, {Jordan}, {Harding}, {Harris}, {Hendry}, {Holder}, {Ivans},
  {Ivezi{\v c}}, {Jester}, {Johnson}, {Kent}, {Kleinman}, {Kniazev},
  {Krzesinski}, {Kron}, {Kuropatkin}, {Lebedeva}, {Lee}, {French Leger},
  {L{\'e}pine}, {Levine}, {Lin}, {Long}, {Loomis}, {Lupton}, {Malanushenko},
  {Malanushenko}, {Margon}, {Martinez-Delgado}, {McGehee}, {Monet}, {Morrison},
  {Munn}, {Neilsen}, {Nitta}, {Norris}, {Oravetz}, {Owen}, {Padmanabhan},
  {Pan}, {Peterson}, {Pier}, {Platson}, {Re Fiorentin}, {Richards}, {Rix},
  {Schlegel}, {Schneider}, {Schreiber}, {Schwope}, {Sibley}, {Simmons},
  {Snedden}, {Allyn Smith}, {Stark}, {Stauffer}, {Steinmetz}, {Stoughton},
  {SubbaRao}, {Szalay}, {Szkody}, {Thakar}, {Sivarani}, {Tucker}, {Uomoto},
  {Vanden Berk}, {Vidrih}, {Wadadekar}, {Watters}, {Wilhelm}, {Wyse}, {Yarger},
  \& {Zucker}}]{yanny-segue}
{Yanny}, B., {Rockosi}, C., {Newberg}, H.~J., {et~al.} 2009, \aj, 137, 4377

\bibitem[{{Yu} {et~al.}(2018){Yu}, {Huber}, {Bedding}, {Stello}, {Hon},
  {Murphy}, \& {Khanna}}]{Yu2018}
{Yu}, J., {Huber}, D., {Bedding}, T.~R., {et~al.} 2018, \apjs, 236, 42

\bibitem[{{Zhao} {et~al.}(2012){Zhao}, {Zhao}, {Chu}, {Jing}, \&
  {Deng}}]{Zhao2012}
{Zhao}, G., {Zhao}, Y.-H., {Chu}, Y.-Q., {Jing}, Y.-P., \& {Deng}, L.-C. 2012,
  Research in Astronomy and Astrophysics, 12, 723

\bibitem[{{Zhong} {et~al.}(2020{\natexlab{a}}){Zhong}, {Chen}, {Wu}, {Li},
  {Bai}, \& {Hou}}]{Zhongjing2020}
{Zhong}, J., {Chen}, L., {Wu}, D., {et~al.} 2020{\natexlab{a}}, \aap, 640, A127

\bibitem[{{Zhong} {et~al.}(2020{\natexlab{b}}){Zhong}, {Chen}, {Wu}, {Li},
  {Bai}, \& {Hou}}]{Zhongjingcatalog}
{Zhong}, J., {Chen}, L., {Wu}, D., {et~al.} 2020{\natexlab{b}}, VizieR Online
  Data Catalog, J/A+A/640/A127

\end{thebibliography}

%\appendix
%  \renewcommand{\appendixname}{Appendix~\Alph{section}}
 %\setcounter{figure}{0}
%\renewcommand{\thefigure}{A\arabic{figure}}

\appendix\renewcommand{\appendixname}{Appendix~\Alph{section}}
 %In this section, we show the comparisons between the stellar parameters coming from  PASTEL and the stellar parameters derived from LAMOST spectra for training (top panels) and testing (bottom panels) stars based on neural-network models
 %we have aThe LAMOST FELLOWSHIP is supported by Special Funding for Advanced Users, budgeted and administrated by Center for Astronomical Mega-Science, Chinese Academy of Sciences (CAMS). Supported by High-performance Computing Platform of Peking University. 
\setcounter{figure}{0}
\renewcommand{\thefigure}{A\arabic{figure}}

%\begin{figure*}
%\centering
%\includegraphics[width=5.5in]{figure/select_rgb_rc_vrard.png}
%\caption{ Same with the Fig.\,\ref{select_rgb_rc_predict} but for the stellar distributions in the plane of $\Delta \nu$ and $\Delta P$ provided by \citet{Vrard2016}.}
%\label{select_rgb_rc_vrard}
%\end{figure*}

%\bsp	% typesetting comment
%\label{lastpage}

\end{document}